\documentclass{aa}  

\usepackage{graphicx}
\usepackage{txfonts}
\usepackage{orcidlink}
\newcommand{\feh}{\mbox{[Fe/H]}\xspace}
\newcommand{\teff}{\ensuremath{T_{\rm eff}}\xspace}
\newcommand{\logg}{\mbox{$\log g_*$}\,}

\newcommand{\kms}{\mbox{km\,s$^{-1}$}}
\newcommand{\ms}{\mbox{m\,s$^{-1}$}}

\begin{document} 

    \title{Discovery of a cold giant planet and mass measurement of a hot super-Earth in the multi-planetary system WASP-132}
    \titlerunning{WASP-132}
    \authorrunning{N. Grieves, et al.}
    \author{Nolan Grieves
    \inst{\ref{inst-geneva}}\fnmsep\thanks{\email{nolangrieves@gmail.com}}\orcidlink{0000-0001-8105-0373}
    \and Fran\c{c}ois Bouchy \inst{\ref{inst-geneva}}
    \and David J. Armstrong \inst{\ref{inst-war},\ref{inst-warexo}}
    \and Babatunde Akinsanmi \inst{\ref{inst-geneva}}
    \and Angelica Psaridi \inst{\ref{inst-geneva}}
    \and Sol{\`e}ne Ulmer-Moll \inst{\ref{inst-geneva},\ref{inst-bern}}
    \and Yolanda G. C. Frensch \inst{\ref{inst-geneva}}
    \and Ravit Helled \inst{\ref{inst-zurich}}
    \and Simon M{\"u}ller \inst{\ref{inst-zurich}}\orcidlink{0000-0002-8278-8377}
    \and Henrik Knierim \inst{\ref{inst-zurich}}
    \and Nuno C. Santos \inst{\ref{inst-porto1},\ref{inst-porto2}}
    \and Vardan Adibekyan \inst{\ref{inst-porto1}}
    \and L{\'e}na Parc\orcidlink{0000-0002-7382-1913} \inst{\ref{inst-geneva}}
    \and Monika Lendl \inst{\ref{inst-geneva}}
    \and Matthew P. Battley \inst{\ref{inst-geneva}}
    \and Nicolas Unger \inst{\ref{inst-geneva}}
    \and Guillaume Chaverot \inst{\ref{inst-geneva}}
    \and Daniel Bayliss \inst{\ref{inst-war},\ref{inst-warexo}}
    \and Xavier Dumusque \inst{\ref{inst-geneva}}
    \and Faith Hawthorn \inst{\ref{inst-war},\ref{inst-warexo}}
    \and Pedro Figueira \inst{\ref{inst-geneva},\ref{inst-porto1}}
    \and Marcelo Aron Fetzner Keniger \inst{\ref{inst-war},\ref{inst-warexo}}
    \and Jorge Lillo-Box \inst{\ref{inst-cab}}
    \and Louise Dyregaard Nielsen \inst{\ref{inst-eso},\ref{inst-geneva}}
    \and Ares Osborn \inst{\ref{inst-war},\ref{inst-warexo}}
    \and Sérgio G. Sousa \inst{\ref{inst-porto1}}
    \and Paul Str{\o}m \inst{\ref{inst-war},\ref{inst-warexo}}
    \orcidlink{https://orcid.org/0000-0002-7823-1090}
    \and St{\'e}phane Udry \inst{\ref{inst-geneva}}
    }
    
   \institute{
    Observatoire de Gen{\`e}ve, Universit{\'e} de Gen{\`e}ve, 51 Chemin Pegasi, 1290 Versoix, Switzerland \label{inst-geneva}
    \and
    Department of Physics, University of Warwick, Gibbet Hill Road, Coventry, UK \label{inst-war}
    \and 
    Center for Exoplanets and Habitability, University of Warwick, Gibbet Hill Road, Coventry, UK \label{inst-warexo}
    \and 
    Physikalisches Institut, University of Bern, Gesellschaftsstrasse 6, 3012 Bern, Switzerland \label{inst-bern}
    \and
    Center for Theoretical Astrophysics \& Cosmology, Institute for Computational Science, University of Zurich, Z\"{u}rich, Switzerland \label{inst-zurich}
    \and 
    Instituto de Astrof\'{i}sica e Ci\^encias do Espa\c{c}o, Universidade do Porto, CauP, Rua das Estrelas, 4150-762 Porto, Portugal \label{inst-porto1}
    \and
    Dep. de F\'{i}sica e Astronomia, Faculdade de Ci\^encias, Universidade do Porto, Rua do Campo Alegre, 4169-007 Porto, Portugal \label{inst-porto2}
    \and
    Centro de Astrobiolog\'ia (CAB), CSIC-INTA, Dep. de Astrof\'isica, ESAC campus, 28692,Villanueva de la Ca\~nada (Madrid), Spain \label{inst-cab}
    \and
    European Southern Observatory, Karl-Schwarzschild-Strasse 3, 85748 Garching, Germany \label{inst-eso}
    } 

   \date{Received 6 October 2023 / Accepted 25 November 2024}

 
  \abstract
   {Hot Jupiters generally do not have nearby planet companions, as they may have cleared out other planets during their inward migration from more distant orbits. This gives evidence that hot Jupiters more often migrate inward via high-eccentricity migration due to dynamical interactions between planets rather than more dynamically cool migration mechanisms through the protoplanetary disk. Here we further refine the unique system of WASP-132 by characterizing the mass of the recently validated 1.0-day period super-Earth WASP-132\,c (TOI-822.02), interior to the 7.1-day period hot Jupiter WASP-132\,b. Additionally, we announce the discovery of a giant planet at a 5-year period (2.7 AU). We also detected a long-term trend in the radial velocity data indicative of another outer companion. Using over nine years of CORALIE radial velocities (RVs) and over two months of highly sampled HARPS RVs, we determined the masses of the planets from smallest to largest orbital period to be M$_{\rm{c}}$\,=\,$6.26^{+1.84}_{-1.83}$\,$M_{\oplus}$, M$_{\rm{b}}$\,=\,$0.428^{+0.015}_{-0.015}$\,$M_{\rm{Jup}}$, and M$_{\rm{d}}\sin{i}$\,=\,$5.16^{+0.52}_{-0.52}$\,$M_{\rm{Jup}}$, respectively. Using TESS and CHEOPS photometry data, we measured the radii of the two inner transiting planets to be R$_{\rm{c}}$\,=\,$1.841^{+0.094}_{-0.093}$\,$R_{\oplus}$ and R$_{\rm{b}}$\,=\,$0.901^{+0.038}_{-0.038}$\,$R_{\rm{Jup}}$. We find a bulk density of $\rho_{\rm{c}}$\,=\,$5.47^{+1.96}_{-1.71}$\,g\,cm$^{-3}$ for WASP-132\,c, which is slightly above the Earth-like composition line on the mass-radius diagram. WASP-132 is a unique multi-planetary system in that both an inner rocky planet and an outer giant planet are in a system with a hot Jupiter. This suggests it migrated via  a rarer dynamically cool mechanism and helps to further our understanding of how hot Jupiter systems form and evolve.}


   \maketitle
%

\section{Introduction}

Some previous studies found that most hot Jupiters do not have nearby planet companions. This was based on non-detections of additional planets in radial velocity (RV) data \citep[e.g.,][]{Wright2009} and photometry data by looking for both additional transit signals \citep[e.g.,][]{Steffen2012,Huang2016,Hord2021} and transit-timing variations (TTVs)\citep[e.g.,][]{Steffen2012,Wang2021,Ivshina2022}. The lack of detected planets may be a result of highly eccentric giant planets clearing out any low-mass inner planets in the system during their migration inward \citep{Mustill2015}. These results suggest that high-eccentricity migration due to dynamical interactions between planets \citep[e.g.,][]{Rasio1996,Weidenschilling1996,LinIda1997} may be a more common migration mechanism for gas giant planets compared to disk-driven migration \citep[e.g.,][]{Goldreich1980,Ward1997,Baruteau2014}.

However, \citet{Wu2023} recently searched for TTVs  across the full four-year \textit{Kepler} \citep{Borucki2010} dataset and found that at least 12$\pm$6\% of hot Jupiters have a nearby companion. Continued surveys and recent advances in the precision of both photometry and RV observations have allowed the detection of some of these companions, including WASP-47 \citep{Becker2015}, Kepler-730 \citep{Zhu2018,Canas2019}, TOI-1130 \citep{Huang2020}, WASP-148 \citep{Hebrard2020}, WASP-132 \citep{Hord2022}, and TOI-2000 \citep{Sha2023}. These companions provide strong evidence that these hot Jupiters migrated via quiescent mechanisms that are dynamically cool and allow nearby planetary companions to remain in the system. 

Giant long-period planets have also been detected in systems with small close-in planets \citep[e.g.,][]{Santos2016}. Several studies have worked to determine the occurrence of long-period ($\gtrsim$\,1 year) giant (M$_{p}$\,$\gtrsim$\,0.3\,M$_{\rm{Jup}}$/95\,M$_{\oplus}$; R$_{p}$\,$\gtrsim$\,0.45\,R$_{\rm{Jup}}$/5\,R$_{\oplus}$) planets, also known as cold Jupiters, in systems that contain close-in ($\lesssim$100 days) small (R$_{p}$\,$\lesssim$\,4\,R$_{\oplus}$; M$_{p}$\,$\lesssim$\,30\,M$_{\oplus}$) planets. These studies have varying results with some suggesting a positive correlation between close-in small planets and cold Jupiters \citep[e.g.,][]{ZhuWu2018,Bryan2019,Herman2019,Rosenthal2022}, while others find a negative correlation \citep[e.g.,][]{Barbato2018,Bonomo2023}. A positive correlation suggests that these two planet populations do not directly compete for solid material \citep[e.g.,][]{ZhuWu2018}. However, multiple theories suggest a negative correlation including that the early formation of a cold Jupiter may prevent the nuclei of smaller planets from migrating inward \citep[e.g.,][]{Izidoro2015}, or that cold Jupiters may reduce the flux of material required to form close-in planets larger than Earth \citep[e.g.,][]{Lambrechts2019}. Recently, \citet{Zhu2024} looked at the metallicity dimension of the super-Earth versus cold Jupiter correlation and found that there is a positive correlation between the two around metal-rich host stars; however, a correlation is unclear for metal-poor host stars due to a limited sample size.

Here we analyze the unique WASP-132 planetary system. WASP-132\,b was first discovered by \citet{Hellier2017} who measured the planet to have a 7.1 day period, 0.41\,$\pm$\,0.03\,$\mathrm{M_{\rm Jup}}$ mass, and 0.87\,$\pm$\,0.03\,$\mathrm{R_{\rm Jup}}$ radius using 23,300 WASP-South \citep{Pollacco2006} observations from 2006 May to 2012 June, 36 1.2-m Euler/CORALIE \citep{Queloz2001} RVs from 2014 March to 2016 March, and TRAPPIST \citep{Jehin2011} photometry data on 2014 May 05. \citet{Hord2022} then later announced the discovery and validation of an inner 1.01 day 1.85\,$\mathrm{R_{\oplus}}$ planet using data from the Transiting Exoplanet Survey Satellite \citep[TESS;][]{Ricker2015}, but did not characterize the mass. Here we use new RV measurements to characterize the mass of the previously discovered super-Earth WASP-132\,c \citep{Hord2022}. We also announce the discovery of a long-period massive giant planet, and we updated all bulk measurements of the system. In Sect. \ref{sec:obs} we present the observations used in this work. In Sect. \ref{sec:analysis} we describe our analysis and results. We discuss our results in Sect. \ref{sec:disc} and finally give our conclusions Sect. \ref{sec:conclusion}.

\section{Observations} \label{sec:obs}

\subsection{Photometry}

\subsubsection{TESS}

As detailed in \citet{Hord2022}, WASP-132 (TOI-822, TIC 127530399) was observed by TESS in Sector 11 from UT 2019 April 23 to May 20 (23.96 days) in CCD 2 of Camera 1 and in Sector 38 from UT 2021 April 29 to May 26 (26.34 days) in CCD 1 of Camera 1. Data for WASP-132 were collected at 2-minute cadence in Sectors 11 and 38 and at 20-second cadence in Sector 38. The data were processed by the TESS Science Processing Operation Center (SPOC) pipeline \citep{Jenkins2016} and were searched by the Transiting Planet Search module \citep[TPS;][]{Jenkins2002,Jenkins2010}. The TPS recovered WASP-132\,c with a period of 1.01153 days with a signal-to-noise ratio (S/N) of 10.6 in the combined data from the two sectors. 

\citet{Hord2022} searched and analyzed the TESS time series using the Presearch Data Conditioning Simple Aperture Photometry (PDC\_SAP) TESS light curves generated by the TESS SPOC pipeline \citep{Smith2012,Stumpe2012,Stumpe2014} at the 2-minute cadence for TESS Sectors 11 and 38. Using the transit least-squares (TLS) search algorithm \citep{Hippke2019}, \citet{Hord2022} recovered the hot Jupiter WASP-132\,b signal as well as the WASP-132\,c signal with a period (1.0119 $\pm$ 0.0032 days), depth, and mid-transit time consistent with the values reported by the SPOC pipeline. \citet{Hord2022} also further established the planetary nature of the WASP-132\,c signal by using the validation tools \texttt{vespa} \citep{Morton2012,Morton2015} and \texttt{TRICERATOPS} \citep{Giacalone2020,Giacalone2021} on the signal and found false-positive probabilities of 9.02\,$\times$\,10$^{-5}$ and 0.0107, respectively. 

WASP-132 was recently observed again by TESS in Sector 65 from UT 2023 May 4 to 2023 May 29. Data for WASP-132 were collected at 2-minute cadence in Sector 65. We run an independent analysis of the TESS data, now including Sector 65 and use the PDC\_SAP TESS light curves generated by the TESS SPOC pipeline at 2-minute cadence for all three sectors. The full TESS light curves are displayed in the Supplementary Material published on \href{https://doi.org/10.5281/zenodo.14271139}{Zenodo}.

\subsubsection{CHEOPS}

We observed three transits of WASP-132\,c with the European Space Agency CHaracterizing ExOPlanets Satellite \citep[CHEOPS;][]{Broeg2013,Benz2021}. CHEOPS is a space telescope with a 32 cm aperture for the purpose of precision follow-up of known planetary systems. We observed transits of WASP-132\,c with CHEOPS on 19 May 2022, 4 June 2022, and 11 June 2022 (Program ID: PR430010, PI: B. Akinsanmi) with an exposure time of 60 seconds. The CHEOPS photometry was processed using the dedicated data reduction pipeline \citep[version 13.1;][]{Hoyer2020}. To identify and correct for systematics affecting CHEOPS light curves, we decorrelate the data against the temporal evolution of the telescope roll angle and the flux of background stars inside the photometric aperture \citep[see e.g.,][]{Bonfanti2021}. We discuss the removal of systematics in our CHEOPS data in Sect. \ref{sec:juliet}. Using only CHEOPS data we obtain a transit depth of 588\,$\pm$\,105 ppm (5.6$\sigma$ detection) for WASP-132\,c. The CHEOPS light curves are displayed in the Supplementary Material published on \href{https://doi.org/10.5281/zenodo.14271139}{Zenodo}. 


\subsection{Spectroscopy}

\subsubsection{CORALIE}

In addition to the 36 RV measurements published by \citet{Hellier2017} and used by \citet{Hord2022}, we obtained 37 more CORALIE RVs for a total of 73 CORALIE observations taken across a time span of over nine years. The CORALIE spectrograph is on the Swiss 1.2 m Euler telescope at La Silla Observatory, Chile \citep{Queloz2001} and has a resolution of $R$\,$\sim$\,60,000. 

The CORALIE spectrograph began observations in June 1998 and went through two significant upgrades in June 2007 and in November 2014 in order to increase the overall efficiency and accuracy of the instrument. The 2007 upgrade replaced CORALIE’s fiber link and cross-disperser optics \citep{Segransan2010}. The 2014 upgrade replaced CORALIE’s fiber link with octagonal fibers \citep{Chazelas2012} and added a Fabry-P\'erot calibration unit \citep{Cersullo2017}. Both interventions on the instrument introduced small offsets between the RV measurements collected before and after each upgrade, depending on such parameters as the spectral type and systemic velocity of the observed star. 

Observations of WASP-132 with CORALIE began in March 2014 and we have 22 observations with the first upgrade (the "COR07" setup) and 51 observations after the November 2014 upgrade with the "COR14" setup. We treat these two datasets as separate instruments to account for any offsets in the measurements. We reduced the spectra with the standard calibration reduction pipeline and computed RVs by cross-correlating with a binary G2 mask \citep{Pepe2002}. We obtained median RV uncertainties of $\sim$21 m\,s$^{-1}$ for our CORALIE RVs. We computed stellar activity indicator data from the CORALIE spectra, including: the cross-correlation function \citep[CCF;][]{Baranne1996} Full-Width at Half-Maximum (FWHM); CCF Bisector velocity span \citep[BIS;][]{Queloz2001}; CCF Contrast, which is the "depth" of the normalized CCF or height of the Gaussian; and the chromospheric indices measured in H$\alpha$ (6562.81 \AA) \citep{Boisse2009}, Na \citep{GomesdaSilva2011}, and Ca~\textsc{ii}~H\&K lines (3968.47 $\AA$ and 3933.66 $\AA$) \citep{Boisse2009}. The CORALIE RVs are displayed in Fig. \ref{fig:RV_2pfit}. 



\subsubsection{HARPS}

After the announcement of an inner planet candidate to the hot Jupiter in the WASP-132 system, we began observing WASP-132 with the HARPS spectrograph \citep{Pepe2002,Mayor2003} to obtain more precise RV measurements and measure the mass of the inner 1-day period planet. HARPS is hosted by the ESO 3.6-m telescope at La Silla Observatory, Chile and has a resolving power of $R$\,$\sim$\,115,000.

We obtained 48 HARPS observations of WASP-132 from 17 January 2022 to 1 April 2022 as part of the NOMADS program (PI Armstrong, 108.21YY.001). The HARPS raw data were reduced according to the standard HARPS data reduction software pipeline\footnote{\href{https://www.ls.eso.org/sci/facilities/lasilla/instruments/harps/doc/index.html}{ls.eso.org/sci/facilities/lasilla/instruments/harps/doc/index.html}} with a spectral-type K5 binary mask. We obtained median RV uncertainties of $\sim$2.2 \ms\,for our HARPS RVs. We computed stellar activity indicators for the HARPS data including CCF FWHM, CCF BIS, contrast of the CCF, H$\alpha$ indices, Na indices, and Ca~\textsc{ii} indices, and the Mt Wilson "S index" S$_{MW}$ \citep{Wilson1978}. The HARPS spectra were also used to derive spectral parameters for WASP-132, as detailed in Sect. \ref{sec:star}. The HARPS RVs are displayed in Fig. \ref{fig:RV_2pfit}. Figure \ref{fig:RV_2pfit} also displays a two-planet Keplerian model plus a linear trend (see Sect. \ref{sec:initrv}) for both the CORALIE and HARPS RVs. 



\section{Analysis and results} \label{sec:analysis}

\begin{table}
\caption{Stellar parameters of WASP-132.}
\resizebox{\columnwidth}{!}{%
\begin{tabular}{lcc}
        \hline\hline
        \noalign{\smallskip}
        Parameter       &       Value   &       Source\\

        \hline
    \noalign{\smallskip}
    \noalign{\smallskip}
    \multicolumn{3}{l}{\underline{Identifying Information}}\\
    \noalign{\smallskip}
    \noalign{\smallskip}
    TESS ID & TIC 127530399 &  TESS \\
    TOI ID & TOI-822 &  TESS \\
    2MASS ID &  2MASS J14302619-4609330  &  2MASS \\
    Gaia ID &  6099012478412247296 & Gaia DR3 \\
    
    \\
    \multicolumn{3}{l}{\underline{Astrometric parameters}}\\
    \noalign{\smallskip}
    \noalign{\smallskip}
    R.A. (J2000, h:m:s)         &        14:30:26.21    &  TICv8    \\
    Dec      (J2000, h:m:s)         &    -46:09:34.26   &  TICv8    \\
    Parallax  (mas) & 8.092 $\pm$ 0.019 & Gaia DR3 \\
    Distance  (pc) & 123.17 $\pm$ 0.57 & \ref{sec:star} \\
    Distance (light years) &  401.7 $\pm$ 1.9 & \ref{sec:star} \\
    \\
    \multicolumn{3}{l}{\underline{Photometric parameters}}\\
    \noalign{\smallskip}
    \noalign{\smallskip}
    B  & 13.142 $\pm$ 0.011    & APASS \\ 
    V & 11.938 $\pm$ 0.046    & APASS \\
    G  & 11.747 $\pm$ 0.020 & Gaia DR3 \\
    B$_P$ & 12.300 $\pm$ 0.020 & Gaia DR3 \\
    R$_P$ & 11.049 $\pm$ 0.020 & Gaia DR3 \\
    J  & 10.257 $\pm$ 0.026 & 2MASS \\
    H  & 9.745 $\pm$ 0.023 & 2MASS \\
    K$_S$  & 9.674 $\pm$ 0.024 & 2MASS \\
    W1 & 9.557 $\pm$ 0.022 & WISE \\
    W2 & 9.638 $\pm$ 0.020 & WISE \\
    W3 & 9.575 $\pm$ 0.040 & WISE \\
    A$_{V}$ & 0.114 $\pm$ 0.11 & Sect. \ref{sec:star} \\
    \\
  \multicolumn{3}{l}{\underline{Bulk parameters}}\\
    \noalign{\smallskip}
    \noalign{\smallskip}
    \teff\,(K) & $4686\pm99$   & Sect. \ref{sec:star}\\
    \feh  & $0.15\pm0.05$ & Sect. \ref{sec:star}\\
    \mbox{[Mg/H]}  & $0.13\pm0.09$ & Sect. \ref{sec:star}\\
    \mbox{[Si/H]}  & $0.27\pm0.11$ & Sect. \ref{sec:star}\\
    \logg (cgs; adopted) & $4.56\pm0.03$   & Sect. \ref{sec:star}\\
    \logg (cgs; HARPS) & $4.55\pm0.29$   & Sect. \ref{sec:star}\\
    Spectral type & K4V & Sect. \ref{sec:star}\\
    Mass ($M_{\odot}$) & $0.789\pm0.039$  & Sect. \ref{sec:star} \\
    Radius ($R_{\odot}$) & $0.758\pm0.032$  & Sect. \ref{sec:star} \\
    $\rho_*$ (g\,cm$^{-3}$) & $2.56^{+0.21}_{-0.18}$  & Sect. \ref{sec:star}\\
    Luminosity ($L_{\odot}$) & $0.266^{+0.017}_{-0.012}$  & Sect. \ref{sec:star} \\
    Age     (Gyrs) & $7.2^{+4.3}_{-4.4}$ & Sect. \ref{sec:star} \\
    log(R'$_{\rm{HK}}$) & $-4.852\pm0.039$ & Sect. \ref{sec:star} \\ 
    v$\sin i$ (km\,s$^{-1}$) & $\sim$3.3 & Sect. \ref{sec:star} \\
    \noalign{\smallskip}
    \hline
    \noalign{\smallskip}
    \end{tabular}}
\label{tab:star_table}
\end{table}   

\subsection{Host star parameters} \label{sec:star}
\citet{Hord2022} compared several methods to determine the host star parameters of WASP-132 including those reported by \citet{Hellier2017}, the TICv8.2 values \citep{Stassun2018,Stassun2019}, an isochrone-based analysis using \texttt{isoclassify} \citep{Huber2017,Berger2020}, as well as two different broadband Spectral Energy Distribution (SED) plus parallax analyses. \citet{Hord2022} found these methods to be consistent and adopted the isochrone analysis as their final parmameters which used spectroscopic \teff and metallicity from \citet{Hellier2017}, Gaia Data Release 2 \citep[DR2;][]{GaiaCollaboration2016,Bailer-Jones2018,GaiaCollaboration2018} parallax and coordinates, the Two Micron All-Sky Survey \citep[2MASS;][]{Skrutskie2006} K$_S$ magnitude, and a photometric extinction estimated from \citet{Bovy2016} as inputs.

The spectroscopic \teff and \feh from \citet{Hellier2017} that \citet{Hord2022} used as inputs for their final model were determined using only 36 CORALIE spectra. We now have 48 higher resolution HARPS spectra and can derive higher precision spectral parameters using a combined HARPS spectrum. The stellar atmospheric parameters ($T_{\mathrm{eff}}$, $\log{g}$, microturbulence and [Fe/H]) were derived using the methodology described in \citet[][]{Sousa-14, Santos-13}. We first measured the equivalent widths (EWs) of 224 Fe I and 35 Fe II lines using the ARES v2 code\footnote{The last version of ARES code (ARES v2) can be downloaded at http://www.astro.up.pt/$\sim$sousasag/ares} \citep{Sousa-15}. Then we used these EWs together with a grid of Kurucz model atmospheres \citep{Kurucz1993} and the radiative transfer code MOOG \citep{Sneden-73} to determine the parameters under assumption of ionization and excitation equilibrium. The abundances of Mg and Si were also derived using the same tools and models as detailed in \citep[e.g.][]{Adibekyan-12, Adibekyan-15}. Although the EWs of the spectral lines were automatically measured with ARES, we performed careful visual inspection of the EWs measurements as only three lines are available for the Mg abundance determination. Using the combined HARPS spectrum we find a \teff = 4686$\pm$99 K, \feh = +0.15$\pm$0.05, and \logg = 4.55$\pm$0.29 (cgs). We find the star to be a K4V spectral type from its \teff value using the updated table from \citet{PecautMamajek2013} \footnote{\url{https://www.pas.rochester.edu/~emamajek/EEM_dwarf_UBVIJHK_colors_Teff.txt}}.

We derived the rotational projected velocity by performing spectral synthesis with MOOG on 36 iron isolated lines, using Gaussian priors for the spectroscopic parameters, and fixing the macroturbulent velocity and limb-darkening coefficient \citep{CostaSilva-20}. The linear limb-darkening coefficient (0.7) was determined using ExoCTK package \citep{Bourque2021} considering the stellar parameters. Given that WASP-132 is a cooler star the \citet{Doyle-14} calibrations are not valid. We assumed a macroturbulent velocity of 2 km\,s$^{-1}$, but the v$\sin i$ value can vary with this assumption. With this method we find a rotational projected velocity of v$\sin i$ = 3.3$\pm$0.6 km\,s$^{-1}$. We note that a v$\sin i$ = 3.3$\pm$0.6 km\,s$^{-1}$ yields a Prot/sini = 11.35$\pm$2.09 days (with the $R_{\star}$ value derived below); however, this is not in agreement with the v$\sin i$ = 0.9$\pm$0.8 km\,s$^{-1}$. \citet{Hellier2017} derived with CORALIE spectra nor with the 33$\pm$3 day signal from a possible rotational modulation in the WASP data. We tested larger values for macroturbulent velocities and found v$\sin i$ values of 2.9, 2.4, and 1.5 km\,s$^{-1}$ for macroturblent velocities of 3, 3.5, and 4 km\,s$^{-1}$, respectively. Given the large variance of v$\sin i$ with the unknown macroturbulent velocity the v$\sin i$ = 3.3$\pm$0.6 km\,s$^{-1}$ can only be taken as an estimate.

To derive our final stellar parameters, including the age of the star, we fit the SED of the star using MESA Isochrones and Stellar Tracks (MIST) \citep{Dotter2016,Choi2016} via the \texttt{IDL} suite \texttt{EXOFASTv2} \citep{Eastman2019}. The stellar parameters are simultaneously constrained by the SED and the MIST isochrones with this method as the SED primarily constrains the stellar radius $R_{\star}$ and effective temperature \teff, while a penalty for straying from the MIST evolutionary tracks ensures that the resulting star is physical in nature (see \citet{Eastman2019}, for more details on the method). 

We use the \teff, \logg, and \feh values along with their uncertainties from the HARPS spectral analysis as priors for our stellar model and the extinction ($A_V$) is limited to the maximum line-of-sight extinction from the Galactic dust maps of \citet{Schlafly2011}. We put a Gaussian prior for parallax from the value and uncertainty in $Gaia$ DR3 \citep{GaiaCollaboration2021}. We note the $Gaia$ DR3 parallax (8.09240 mas) was corrected by subtracting $-$0.02705 mas according to the \citet{Lindegren2021} prescription. For the SED fit we use photometry from APASS DR9 $BV$ \citep{Henden2016}; $Gaia$ DR3 G, B$_P$, and R$_P$ \citep{GaiaCollaboration2021}; 2MASS J, H, and K$_S$ \citep{Skrutskie2006}; and ALL-WISE W1, W2, and W3 \citep{Wright2010}, which are presented in Table \ref{tab:star_table}. 

We also use the HARPS spectra to estimate the log(R'$_{\rm{HK}}$) of WASP-132 with the method descirbed by \citet{GomesdaSilva2021} using the bolometric correction by \citet{SuarezMascareno2015} and find a log(R'$_{\rm{HK}}$) = $-$4.852 $\pm$ 0.039, which suggests that WASP-132 is only a moderately chromospherically active or inactive star \citep[e.g.,][]{Henry1996}. From this log(R'$_{\rm{HK}}$) value we obtain a chromospheric rotational period via the relation by \citet{Noyes1984} and \citet{MamajekHillenbrand2008} to be 44\,$\pm$\,8 days which is closer to the 33$\pm$3 day estimate by \citet{Hellier2017} from photometry variation than our estimate from v$\sin i$. 

We present our final stellar parameters in Table \ref{tab:star_table}. The \teff and \feh values are from the HARPS spectral analysis, while the \logg, $A_V$, stellar mass, age, density, luminosity, and distance are outputs of the \texttt{EXOFASTv2} fit. Overall, we find our stellar parameters including mass and radius to be close in value and within uncertainties of those presented by \citet{Hord2022}, as well as a close agreement between the HARPS spectroscopic derived \logg and that derived from \texttt{EXOFASTv2}.

\subsection{Initial RV analysis} \label{sec:rv}

\begin{figure}
  \centering
  \includegraphics[width=0.49\textwidth]{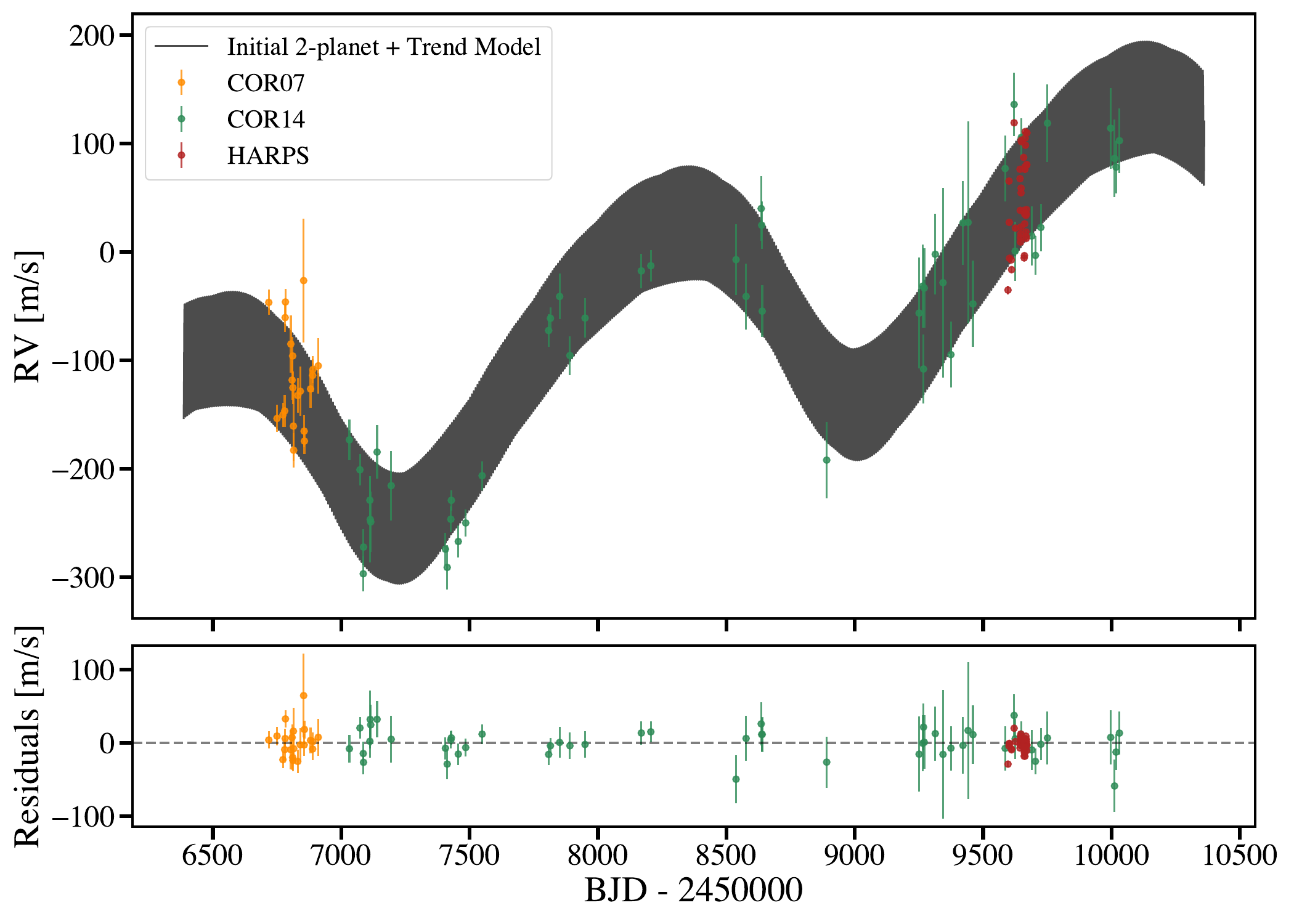}
    \includegraphics[width=0.49\textwidth]{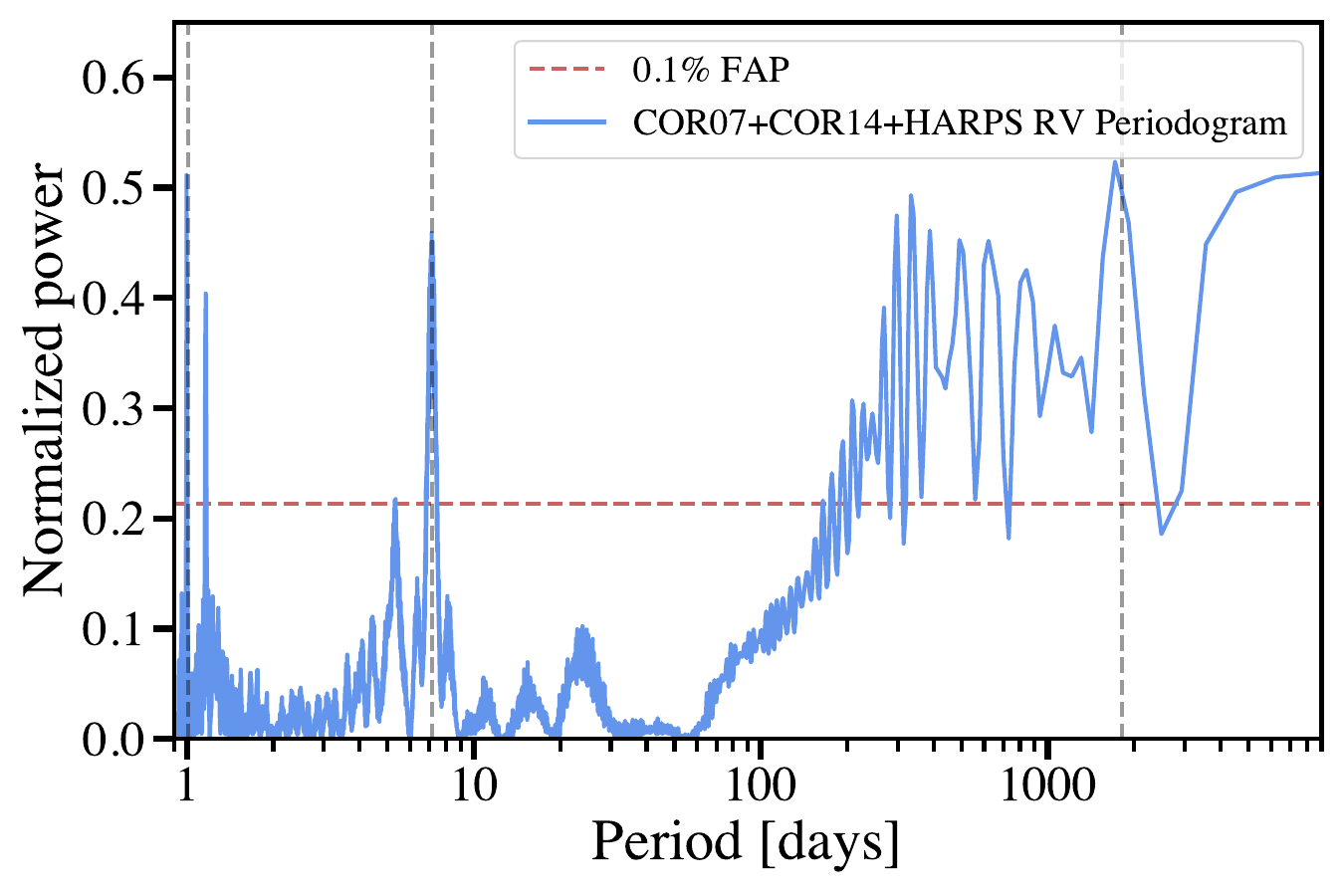}
 \caption{WASP-132 RVs and periodogram. Top: WASP-132 CORALIE and HARPS RVs with a two-planet (excluding the 1-day planet) Keplerian model plus a linear trend as described in Sect. \ref{sec:initrv}. Bottom: GLS periodogram of the original CORALIE and HARPS RVs. The vertical gray dashed lines display the planet periods presented in Sect. \ref{sec:juliet}.}
  \label{fig:RV_2pfit}
\end{figure}

\begin{figure}
  \centering
  \includegraphics[width=0.49\textwidth]{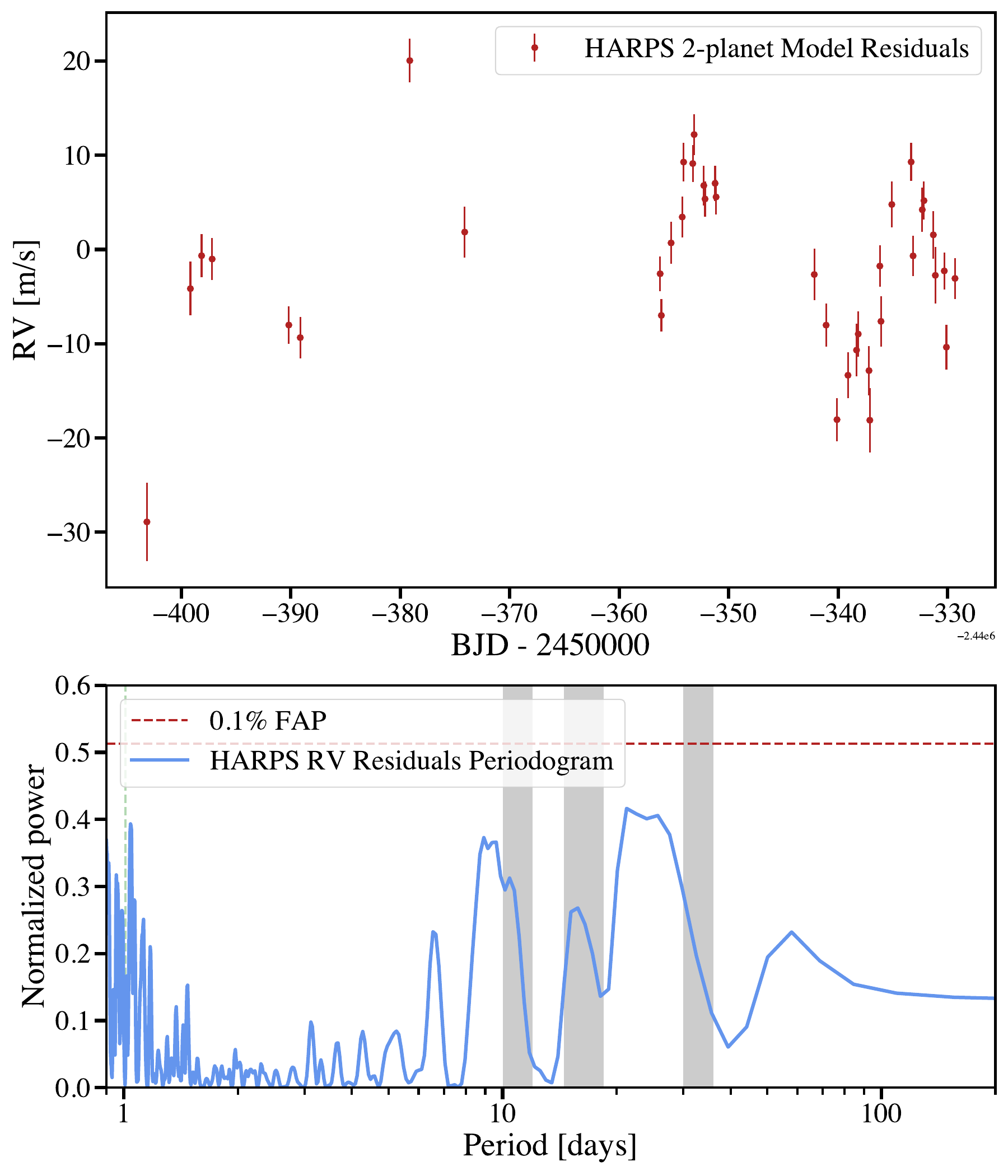}
    \includegraphics[width=0.49\textwidth]{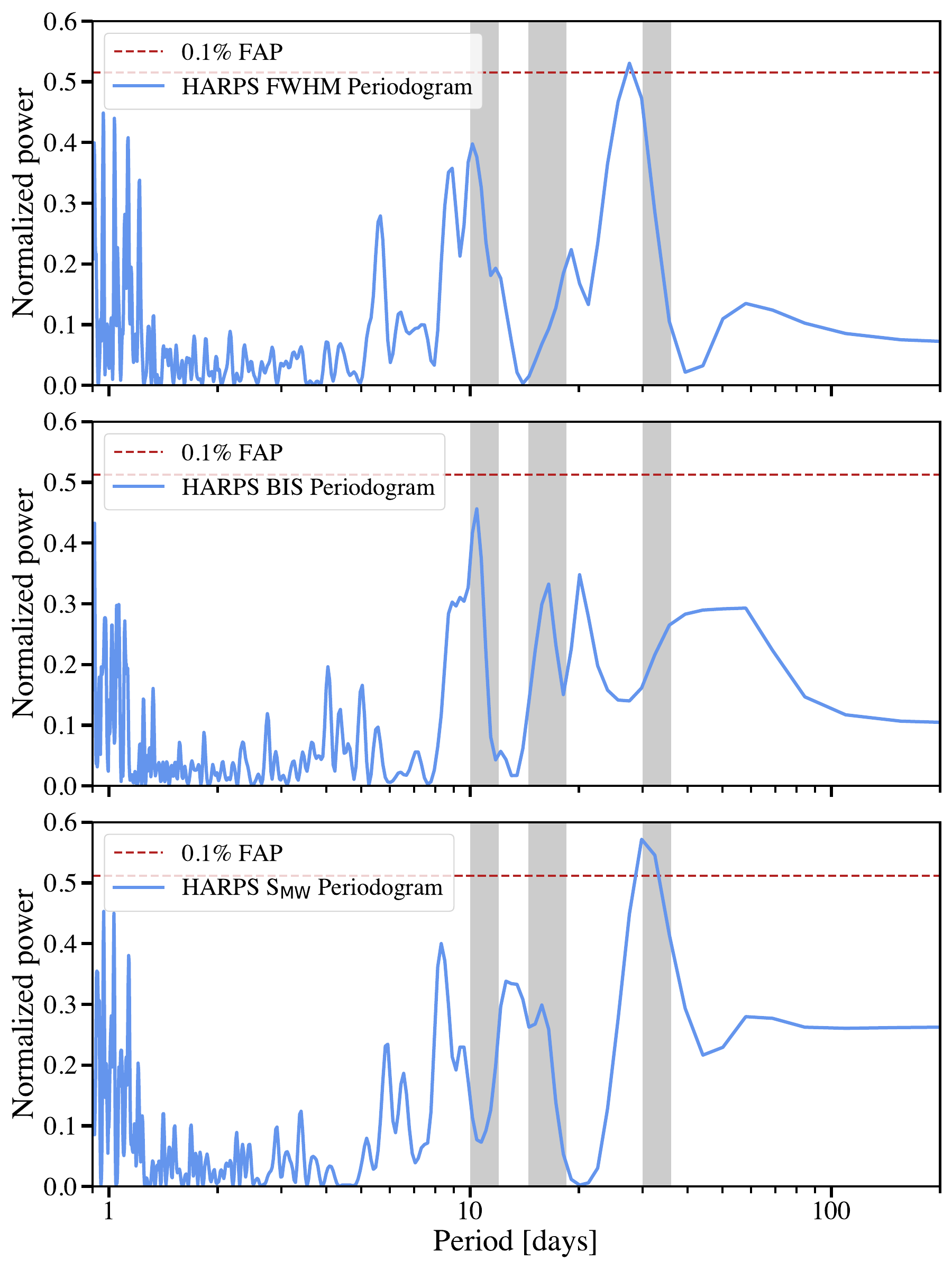}
 \caption{HARPS RV residuals, periodogram, and activity periodograms. Top: HARPS RV residuals after fitting CORALIE and HARPS data with a two-planet plus trend model and its corresponding and GLS periodogram. Bottom: GLS periodograms for the HARPS data activity indicators CCF FWHM, CCF BIS, and Mt Wilson S index S$_{MW}$ from top to bottom respectively. The gray vertical bands show the possible rotation period P$_{\rm{rot}}$\,=\,33\,$\pm$\,3\,days and harmonics P$_{\rm{rot}}$/2 and P$_{\rm{rot}}$/3.}
  \label{fig:harps_resid_GLS}
\end{figure}

\begin{figure}
  \centering
  \includegraphics[width=0.49\textwidth]{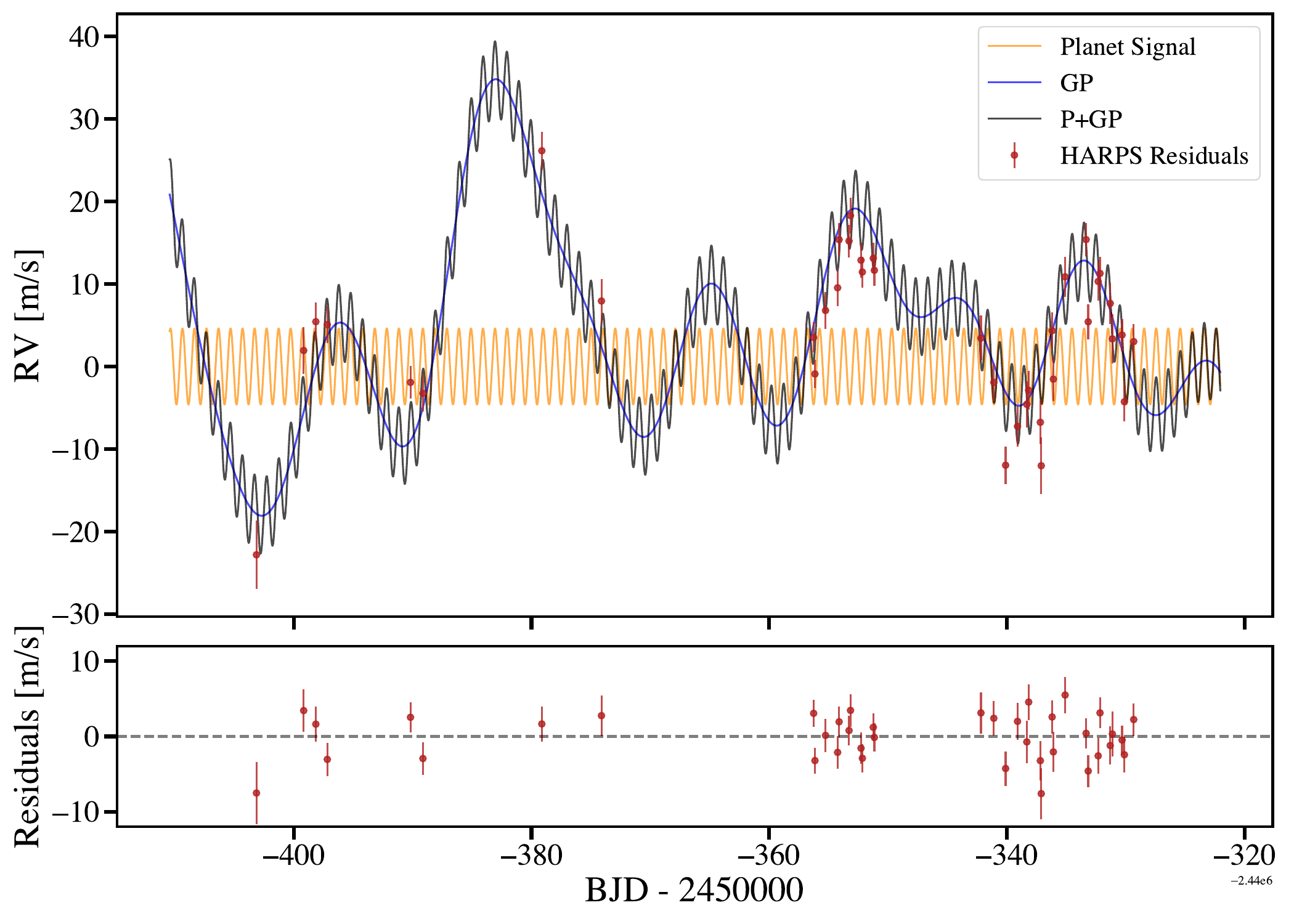}
  \includegraphics[width=0.47\textwidth]{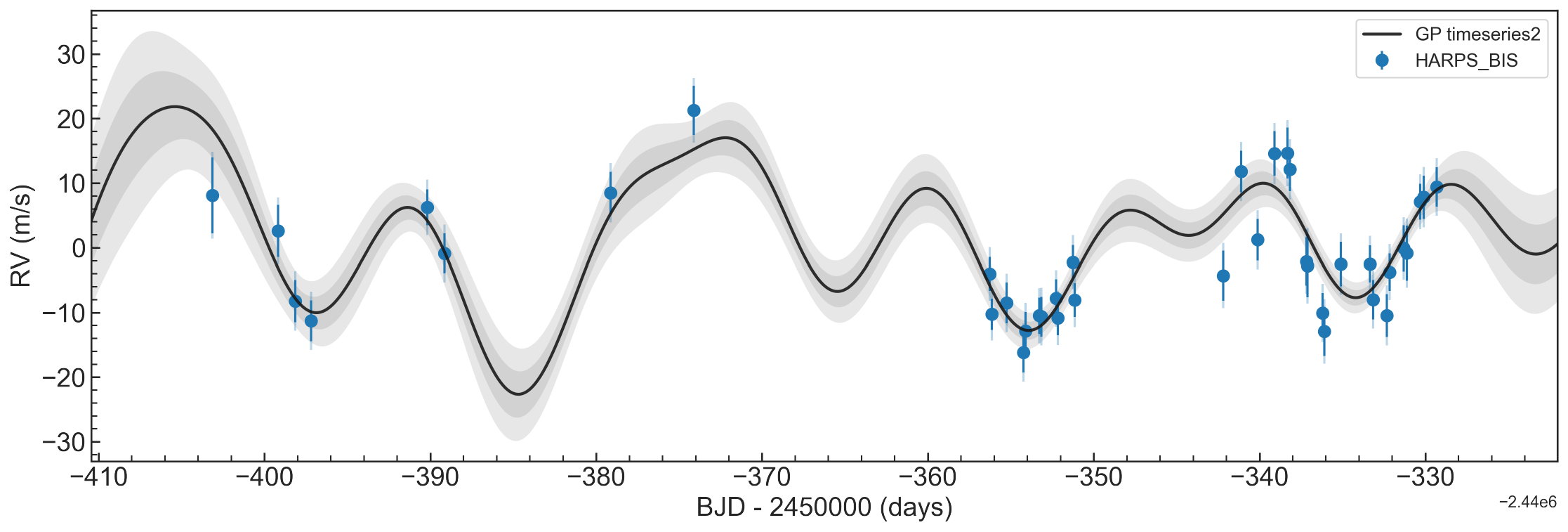}
    \includegraphics[width=0.49\textwidth]{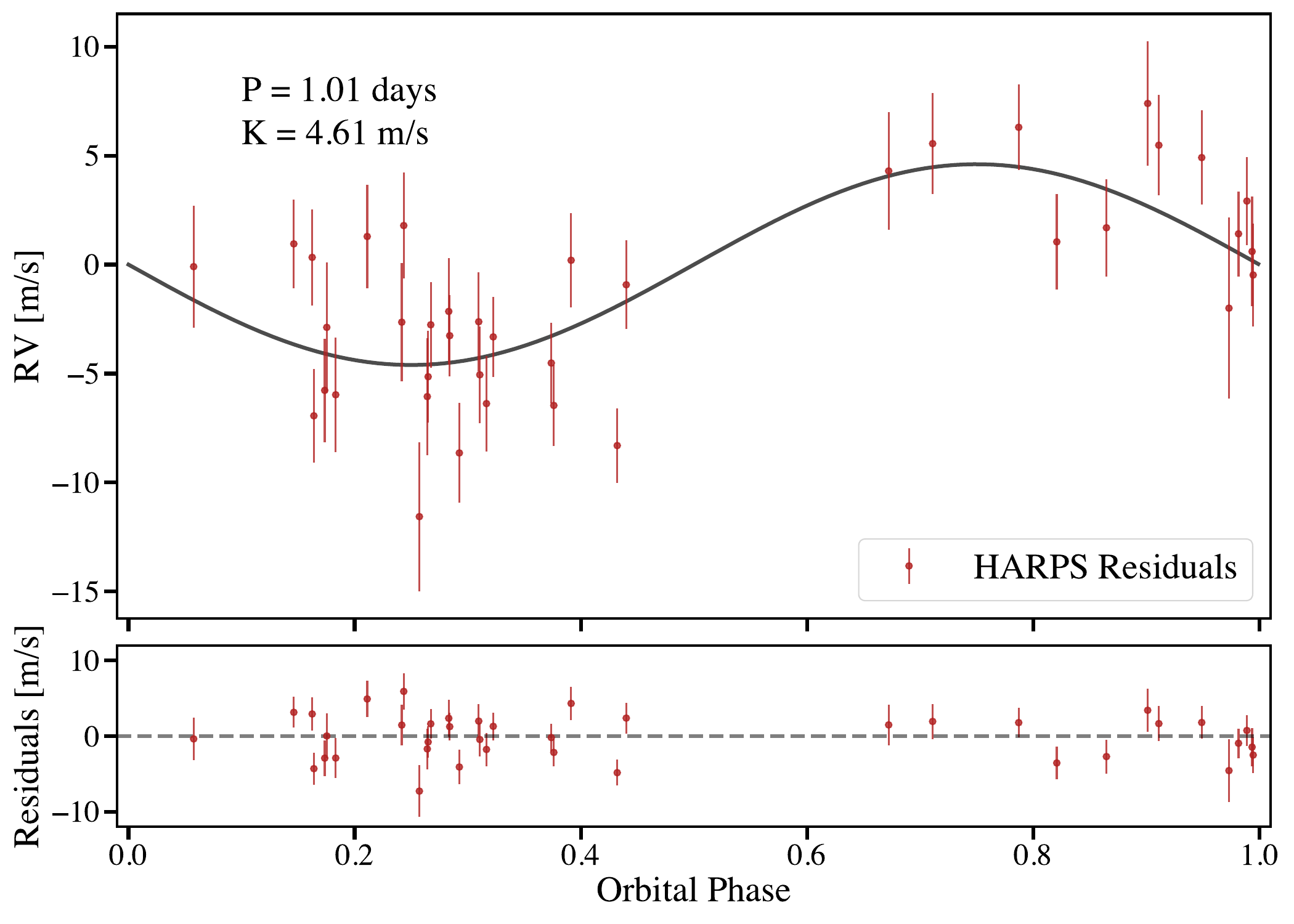}
 \caption{HARPS RV residuals Keplerian and GP model fit. Top: HARPS RV residuals after fitting CORALIE and HARPS data with a two-planet plus trend model described in Sect. \ref{sec:initrv} with a one-planet Keplerian plus GP model constrained by the HARPS RV and BIS. Middle: GP fit to the HARPS BIS data. Bottom: HARPS RV residuals from the top phased to the 1-day planet model.}
  \label{fig:harps_resid_GP}
\end{figure}

\begin{figure}
  \centering
  \includegraphics[width=0.49\textwidth]{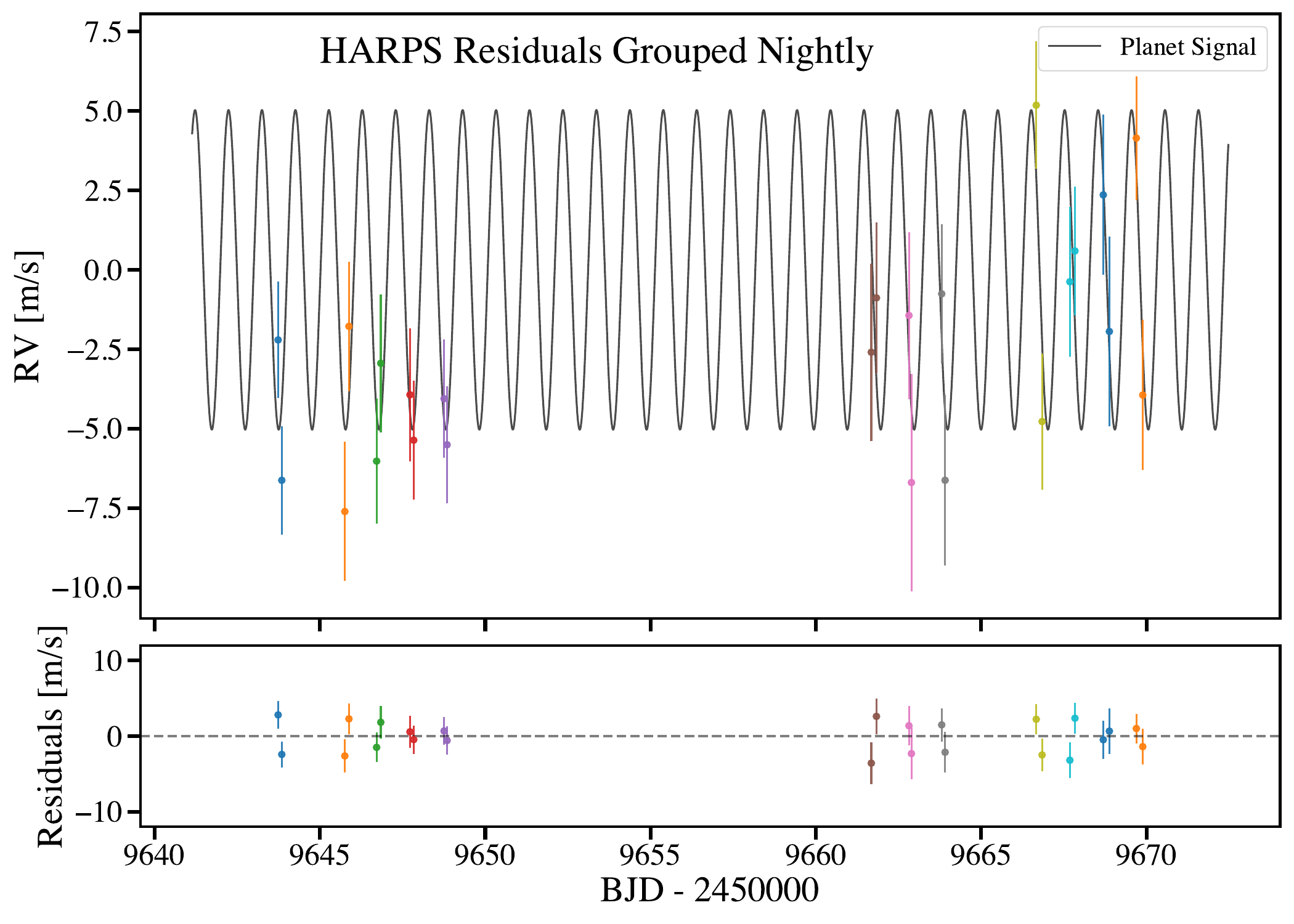}
    \includegraphics[width=0.49\textwidth]{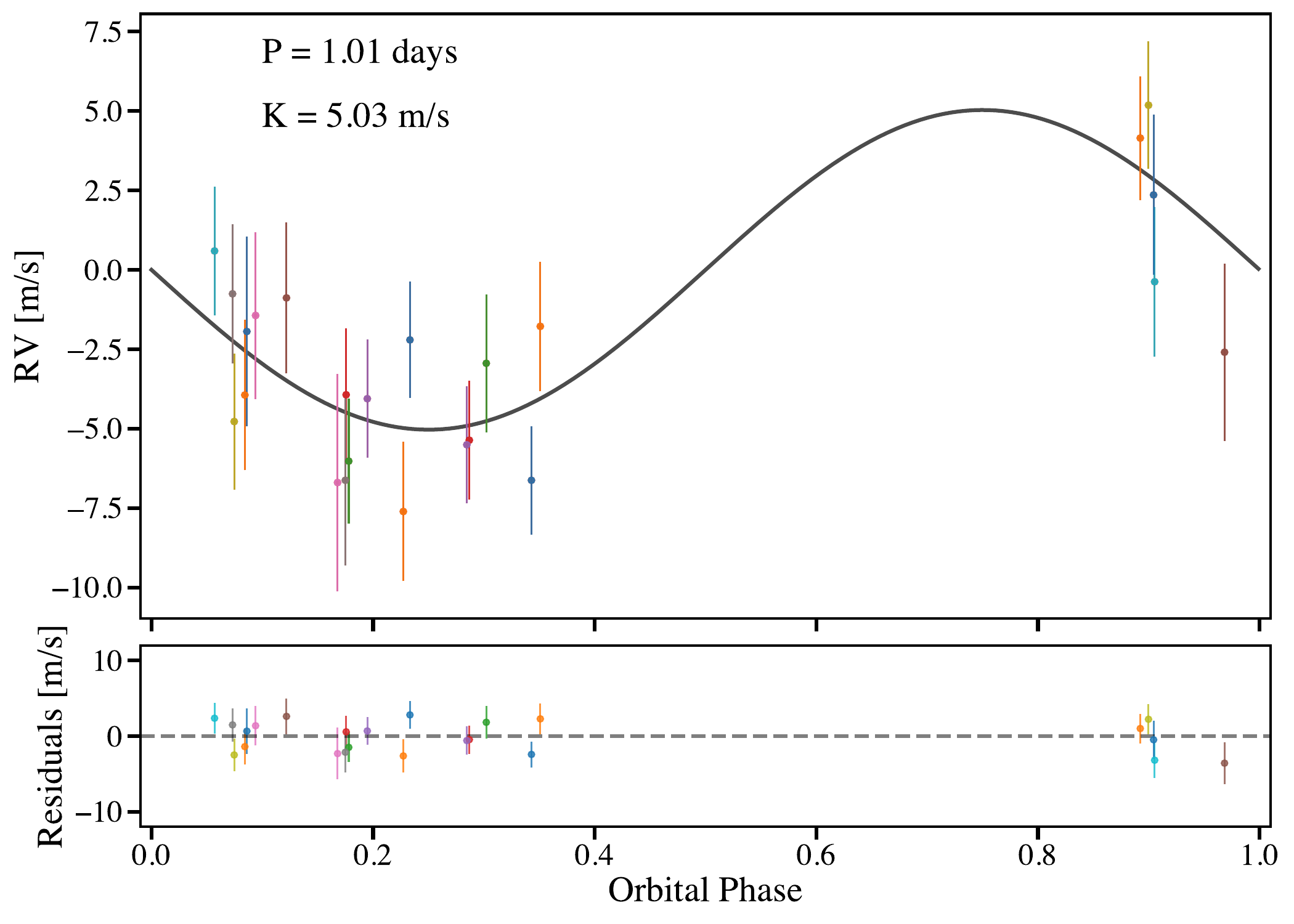}
 \caption{HARPS RV residuals after fitting CORALIE and HARPS data with a two-planet plus trend model described in Sect. \ref{sec:initrv}. Here we only include HARPS data that had two observations in one night for a total of 12 nights. We then fit a single Keplerian model with each night treated as different instrument with its own offset. Each night is shown with different colors. We show the phased RV plot in the bottom.}
  \label{fig:2pfit_resid_bin}
\end{figure}

\subsubsection{Rossiter McLaughlin effect} \label{sec:rm}

The Rossiter-McLaughlin effect \citep{Rossiter1924,McLaughlin1924} occurs during a planetary transit and impacts RVs from a typical Keplerian motion. We estimated the expected RV amplitude of the Rossiter-McLaughlin effect for both transiting planets with the classical method (Eq. 40 from \citealt{Winn2010}) using our v$\sin i$ and stellar radius measurements along with our derived values of the planets' radii and impact parameters, which we later describe in Sect. \ref{sec:juliet}. We find Rossiter-McLaughlin effect RV amplitudes of $\sim$1.6\,m\,s$^{-1}$ for planet c and $\sim$47\,m\,s$^{-1}$ for planet b, assuming $\lambda$\,=\,0 (no obliquity). Thus it is necessary to remove RV data that occurred during transits to not affect the RV model parameters. Using the period and transit duration that we later describe in Sect. \ref{sec:juliet}, we find three CORALIE RVs and four HARPS RVs that occur during a transit of planet c. We removed the four HARPS RVs from our analysis, but given the relatively small impact the planet c Rossiter-McLaughlin effect will have on CORALIE data, due to the CORALIE RV uncertainties, we did not remove the three CORALIE RVs. We also find one CORALIE RV and five HARPS RVs that are during a transit of planet b and we remove all of these data from our analysis. We note that high resolution ESPRESSO \citep{Pepe2021} observations of the Rossiter-McLaughlin signal for both planets were obtained and are in preparation for publication by a separate team.  

\subsubsection{HARPS linear trend plus two-planet residuals analysis} \label{sec:initrv}

Our CORALIE and HARPS RVs clearly show the signal of the 7-day hot Jupiter and a longer period companion around 1800 days as seen in the generalized Lomb-Scargle (GLS) periodogram \citep{Lomb1976,Scargle1982,Zechmeister2009} implemented in the \texttt{astropy} \citep{Astropy2013,Astropy2018} Python package of all the WASP-132 RVs in Fig. \ref{fig:RV_2pfit}. The \texttt{astropy} periodogram package can calculate False Alarm Probabilities (FAP) using various methods and we use the default \citet{Baluev2008} method. We accounted for RV offsets between the instruments using the best-fitting systemic velocities, as derived from the COR07, COR14, and HARPS datasets (see Table \ref{tab:pyaneti}). Notably, we see the 7-day and long-period companion as well as one-day signals above the 0.1\% FAP level in the periodogram. We also detected that a long-term RV trend is within the data. 

In order to detect the much smaller signal from the 1-day period super-Earth planet we first fit the longer period signals and analyzed the residuals. For our initial RV analysis, we used the software suite \texttt{pyaneti}\footnote{https://github.com/oscaribv/pyaneti} \citep{Barragan2019,Barragan2022}, which couples a Bayesian framework with a Markov chain Monte Carlo (MCMC) sampling to produce posterior distributions of the fitted parameters. We first fit the RVs with two Keplerians for the hot Jupiter and long-period outer companion as well as a linear trend. We note that we fit the data without a linear trend and the linear trend model is preferred with a difference in Bayesian Information Criterion (BIC) of 134. We put Gaussian priors on the period of the hot Jupiter based on our transit analysis presented in Sect. \ref{sec:juliet}. With our initial two-planet fit we find the outer companion to have a period of $\sim$1792 days and RV semi-amplitude $K_{\rm{d}}$ = 105.6\,$\pm$\,7.4\,m\,s$^{-1}$ and a linear trend of 23.5$^{+2.4}_{-2.6}$\,m\,s$^{-1}$\,year$^{-1}$. After removing these signals we analyzed the HARPS RVs which have the necessary precision to detect the smaller 1-day planet. 

Noticeably the HARPS RV residuals of the 2-planet plus trend fit have variation much larger than the expected signal from the 1-day period super-Earth planet and we find strong signals at $\sim$21-28 and $\sim$8-11 days as displayed in Fig. \ref{fig:harps_resid_GLS}. It is well known that magnetically active regions (spots and plages) can induce periodic and quasi-periodic Doppler signals at the stellar rotation frequency and its harmonics \citep[e.g.,][]{Boisse2011}. Our log(R'$_{\rm{HK}}$) = $-$4.852\,$\pm$\,0.039 value suggests that the star may be magnetically active and additionally \citet{Hellier2017} found a possible rotational modulation with a period of 33\,$\pm$\,3 days and an amplitude of 0.4 mmag in three out of five seasons of WASP data. We expect these signals in the HARPS data are due to magnetic activity and rotation of the star. We also detect significant signals in the various HARPS activity indicators. We display GLS periodograms in the period domain of the CCF FWHM, CCF BIS, and Mt Wilson S index S$_{MW}$ in Fig. \ref{fig:harps_resid_GLS}. We present a more detailed view of the HARPS activity indicator data in Sect. \ref{sec:rv_sig_act}.

In order to account for this activity signal we applied a multidimensional Gaussian process (GP) approach, the \texttt{pyaneti} implementation of which is as described in \citet{Rajpaul2015}. We used the quasi-periodic (QP) kernel, as defined by \citet{Roberts2013} and described in \citet{Barragan2022}, and placed an uninformative prior with a range of 5 to 45 days for the period of the QP kernel. We tested combinations of available activity indicators and in all cases the period of QP kernel was found to be at P$_{GP}$ $\sim$30-32 days, which we take to be the rotation period of the star. We found that a two-dimensional GP fit with the RV and BIS activity indicators gave the best results and obtained the lowest BIC value and smaller HARPS RV jitter value when we allow an RV jitter to be fit. With this fit we found a RV semi-amplitude of $K$\,=\,4.61$_{-1.75}^{+1.79}$\,m\,s$^{-1}$ corresponding to a mass of M$_{p}$\,=\,6.2$_{-2.3}^{+2.4}$\,M$_{\oplus}$ for the 1-day planet and a P$_{GP}$\,=\,31.22$^{+0.99}_{-1.27}$ days. We show the multidimensional GP fit with BIS and the HARPS RV residuals in Fig. \ref{fig:harps_resid_GP}.

For the previous fits we fixed the eccentricity of the 1-day period planet to 0. We also tested for eccentricity by fitting for $e$cos$\omega$ and $e$sin$\omega$. The fit finds an eccentricity of e = 0.20$_{-0.14}^{+0.24}$. However, the BIC is actually higher by 3 for the fit with the eccentricity. Given the large errors on the eccentricity, the smaller BIC of the fixed eccentricity, and the physical likelihood that the planet is tidally locked we fix the eccentricity to 0. Additionally given the complexity of the activity and small planet signal we conclude we that we cannot measure the eccentricity of the 1-day planet with the current RV dataset.

Given the short period of the 1-day planet signal and longer rotation period of the star, we also tested fitting only HARPS RVs that had multiple observations taken on a given night and fit each night as its own offset in order to constrain the RV semi-amplitude of only the planet signal, similar to the analysis presented in \citet{Hatzes2011} to obtain the mass of CoRoT-7b \citep{Leger2009,Queloz2009}. After removing RVs during transits, we had 12 different nights of HARPS data with two observations per night. Allowing each night to have its own RV offset and then fitting for a single Keplerian we found an RV semi-amplitude of $K$\,=\,5.03$_{-1.40}^{+1.40}$\,m\,s$^{-1}$ corresponding to a mass of M$_{p}$\,=\,6.7$_{-1.9}^{+1.9}$\,M$_{\oplus}$, which is consistent with our Keplerian plus GP model of the residuals. 

\begin{table}
\centering
\caption{\texttt{Pyaneti} initial RV fit parameters of WASP-132.}
\begin{tabular}{lc}
        \hline\hline
        \noalign{\smallskip}
        Parameter       &       Value   \\
        \hline
    \noalign{\smallskip}
    \noalign{\smallskip}
    T0$_{c}$ [BJD\_TDB] & $2458597.5761_{-0.0023}^{+0.0025}$ \\
    P$_{c}$ [days] & $1.0115337 \pm 0.0000050$ \\
    ecc$_{c}$  & 0 (fixed) \\
    $\omega_{c}$ [deg] & $178_{-122}^{+122}$ \\
    K$_{c}$[${\rm m\,s^{-1}}$] & $4.60_{-1.28}^{+1.37}$ \\
    M$_{c}$ [$M_{\oplus}$] & $6.20_{-1.72}^{+1.84}$ \\
\hline
    T0$_{b}$ [BJD\_TDB] & $2459337.60557_{-0.00010}^{+0.00010}$ \\
    P$_{b}$ [days] & $7.1335139_{-0.0000039}^{+0.0000037}$ \\
    ecc$_{b}$ & $0.0119_{-0.0082}^{+0.0110}$ \\
    $\omega_{b}$ [deg] & $ 66.7_{-50.5}^{+246.7} $ \\
    K$_{b}$[${\rm m\,s^{-1}}$] & $ 52.85_{-1.28}^{+1.34} $ \\
    M$_{b}$ [$M_{\rm{Jup}}$] & $ 0.427_{-0.018}^{+0.018} $ \\
\hline
    T0$_{d}$ [BJD\_TDB] & $ 2455059.5_{-88.3}^{+95.0} $ \\
    P$_{d}$ [days] &  $ 1811.8_{-44.4}^{+42.6} $ \\
    ecc$_{d}$ & $ 0.100_{-0.068}^{+0.078} $ \\
    $\omega_{d}$ [deg] & $ 135.2_{-37.1}^{+30.1} $ \\
    K$_{d}$[${\rm m\,s^{-1}}$] & $ 103.6_{-8.8}^{+9.6} $ \\ 
    M$_{d}$sin$i$ [$M_{\rm{Jup}}$] & $ 5.29_{-0.46}^{+0.48} $ \\
\hline
    COR07 RV [${\rm km\,s^{-1}}$] & $ 31.1188_{-0.0328}^{+0.0252} $ \\
    COR14 RV [${\rm km\,s^{-1}}$] & $ 31.1313_{-0.0057}^{+0.0055} $ \\
    HARPS RV [${\rm km\,s^{-1}}$] & $ 31.1854_{-0.0097}^{+0.0093} $ \\ 
    jitter$_{\rm{COR07}}$ [${\rm m\,s^{-1}}$] & $ 1.95_{-1.57}^{+4.29} $ \\
    jitter$_{\rm{COR14}}$ [${\rm m\,s^{-1}}$] & $ 1.68_{-1.33}^{+3.33} $ \\
    jitter$_{\rm{HARPS}}$ [${\rm m\,s^{-1}}$] & $ 1.62_{-0.90}^{+0.79} $ \\
    linear trend [${\rm m\,s^{-1}\,d^{-1}}$] & $ 0.0645_{-0.0070}^{+0.0066} $ \\

\hline
GP A0 [${\rm km\,s^{-1}}$] & $ 0.0077_{-0.0051}^{+0.0047} $ \\
GP A1 [${\rm km\,s^{-1}}$] & $ 0.052_{-0.031}^{+0.043} $ \\
GP $\lambda_{e}$ [days] & $ 62.5_{-33.9}^{+26.3} $ \\
GP $\lambda_{p}$ & $ 0.59_{-0.18}^{+0.27} $ \\
GP Period [days] & $ 31.42_{-0.59}^{+0.37} $ \\

    \noalign{\smallskip}
    \hline
    \noalign{\smallskip}
    \end{tabular}
\label{tab:pyaneti}
\end{table}   

\begin{figure}
  \centering
  \includegraphics[width=0.49\textwidth]{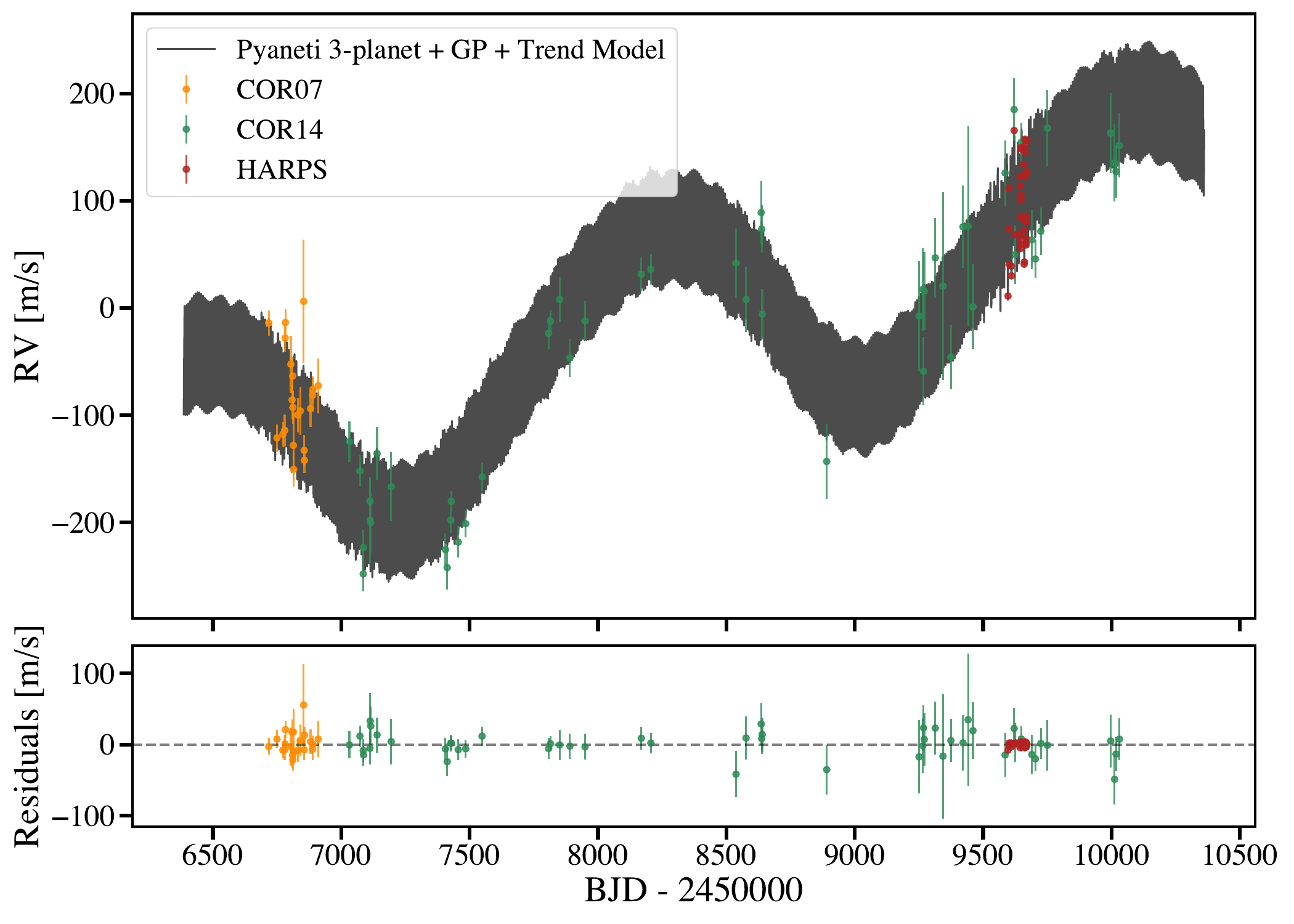}
  \includegraphics[width=0.4\textwidth]{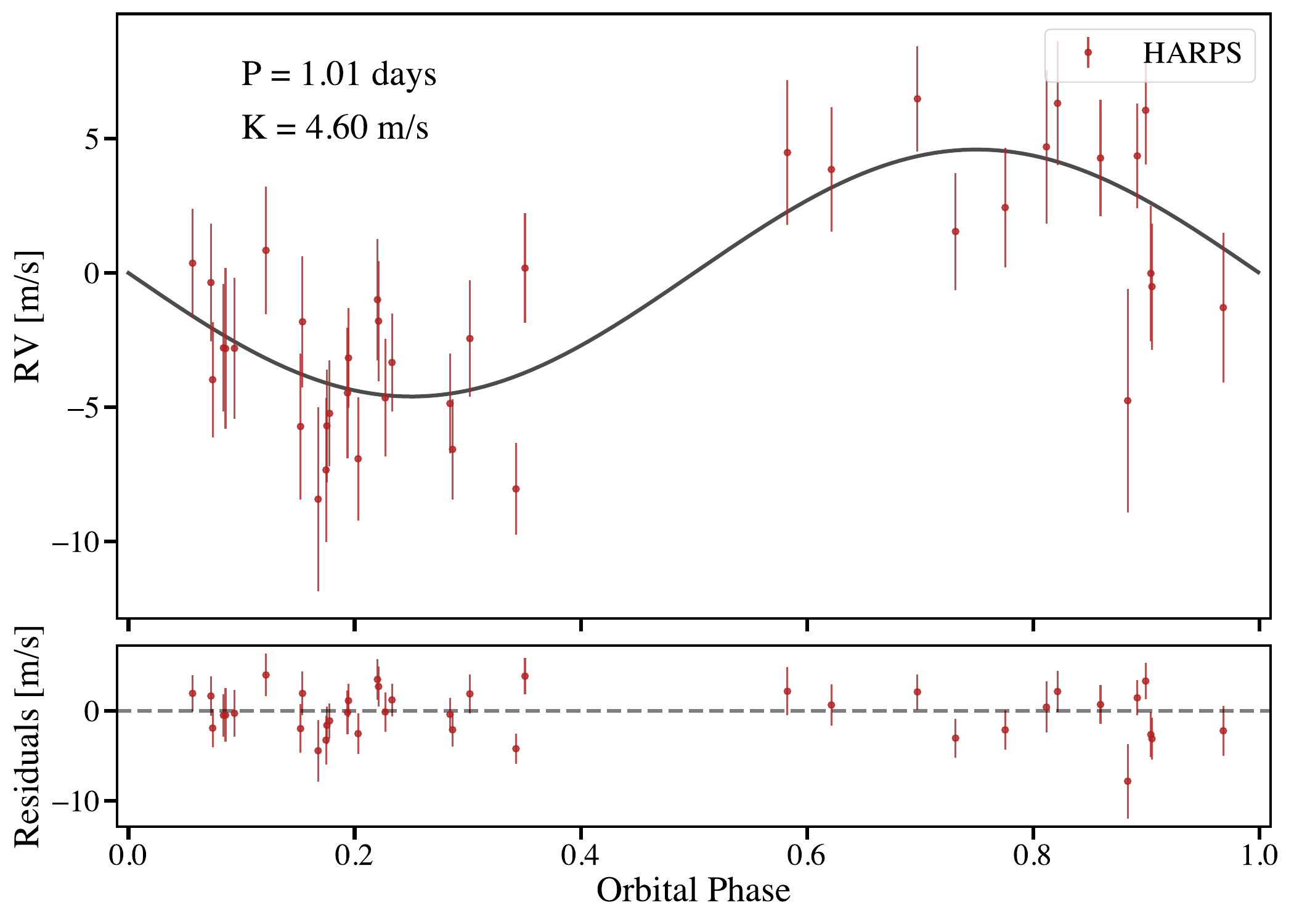}
  \includegraphics[width=0.4\textwidth]{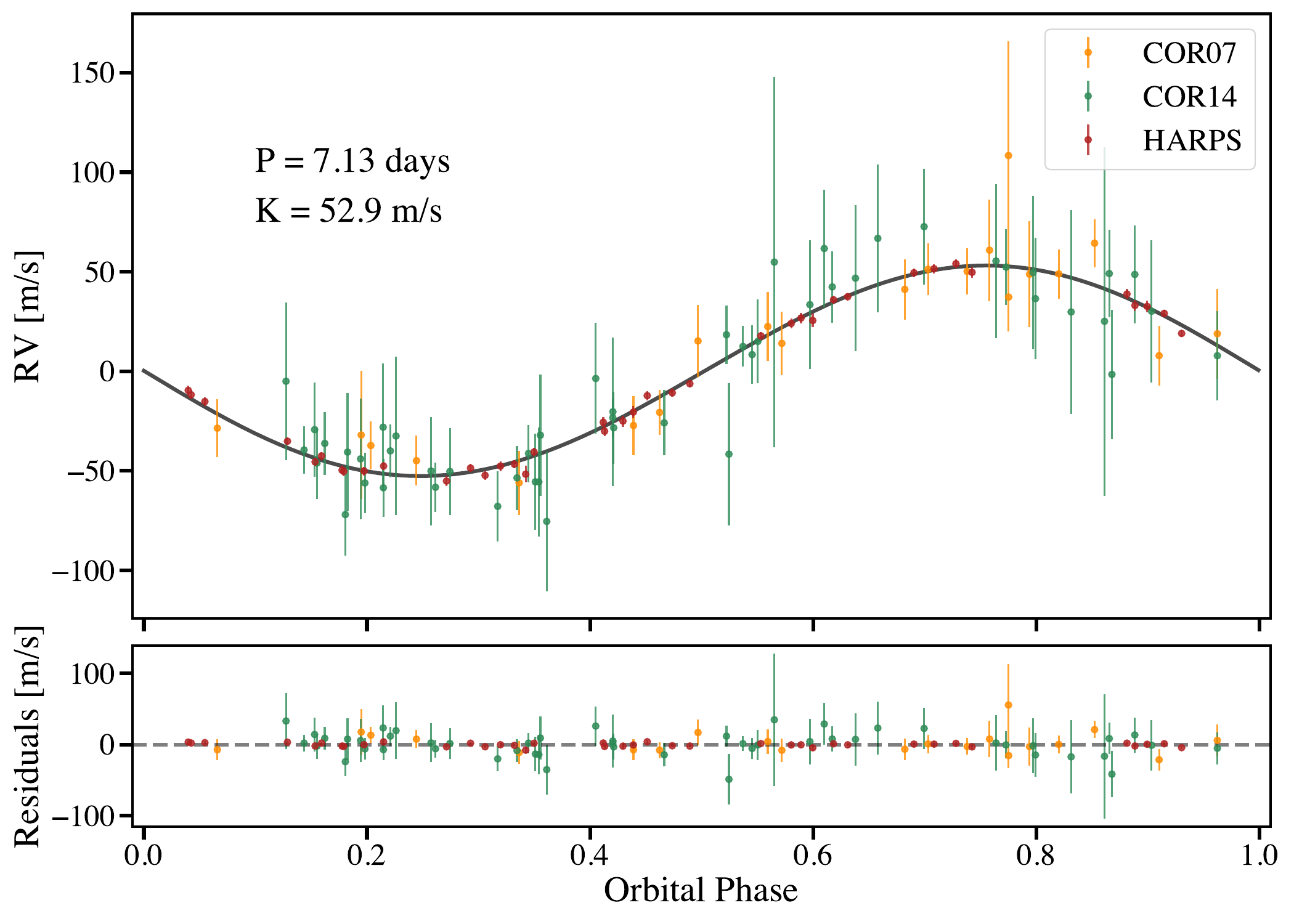}
  \includegraphics[width=0.4\textwidth]{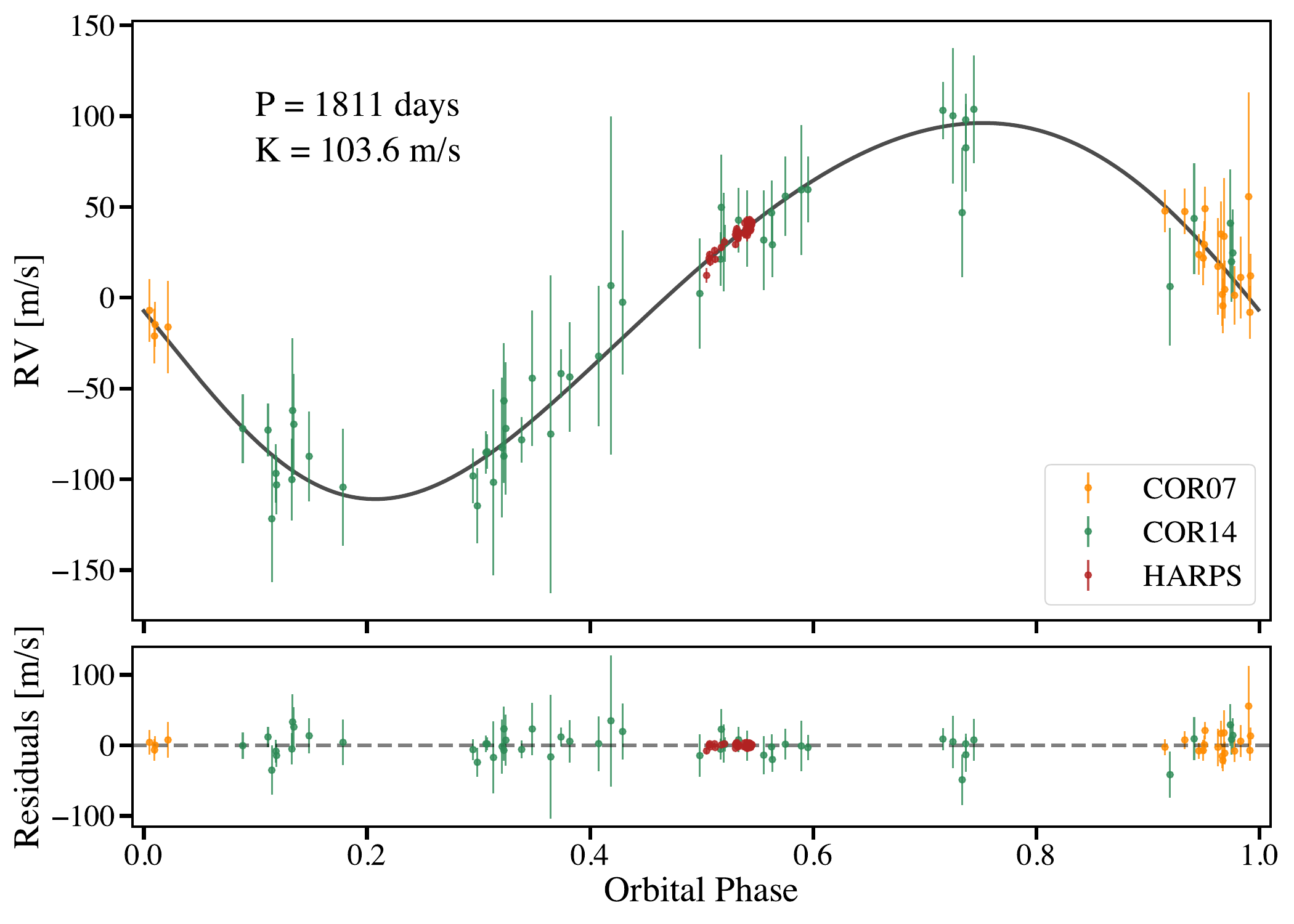}
 \caption{\texttt{pyaneti} fit to RV data including three Keplerian models, a linear trend, and a GP described in Sect. \ref{sec:initrv}. The lower three plots show the phased RV curves of the 1-day, 7-day, and $\sim$1811-day outer planet respectively.}
  \label{fig:pyanet_full}
\end{figure}

\subsubsection{Linear trend plus three-planet RV-only model} \label{sec:fullrv}

Finally for our RV-only analysis we perform a three-planet fit plus a linear trend and a QP GP kernel. We only use a one-dimensional GP as \texttt{pyaneti} cannot fit activity indicators for one instrument and not others, and we do not find a good fit when including the CORALIE activity indicators which are essentially noise when trying to fit the activity signal. We put Gaussian priors on the periods and ephemerides of the two inner planets based on our transit fit in Sect. \ref{sec:juliet}. For the outer planet period we used uniform priors between 1700 and 1900 days. We also put uniform priors on the QP GP period between 30 and 32 days based on our results from the two-dimensional GP modeling with the HARPS residuals. We fixed the 1-day planet's eccentricity to 0. All other parameters were given uniform priors. We show our full fit to the RV data in Fig. \ref{fig:pyanet_full} and present the fit parameters including our mass measurements in Table \ref{tab:pyaneti}, which is in agreement with our previous analyses. Table \ref{tab:pyaneti} presents the QP GP hyper-parameters including A0 and A1, which are the coefficients of the GP and its derivative, respectively. A0 and A1 are amplitudes that act as scale factors, determining the typical deviation from the mean function. $\lambda_{e}$ is the long-term evolution time-scale (the lifetime of active regions). $\lambda_{p}$ is the inverse of the harmonic complexity (how complex variations are inside each period, which is related to distribution of active regions). $\lambda_{e}$ is the long-term evolution time-scale (the lifetime of active regions). In Table \ref{tab:pyaneti} T0$_{d}$ refers to the time of inferior conjunction.

\begin{figure}
  \centering
\includegraphics[width=0.49\textwidth]{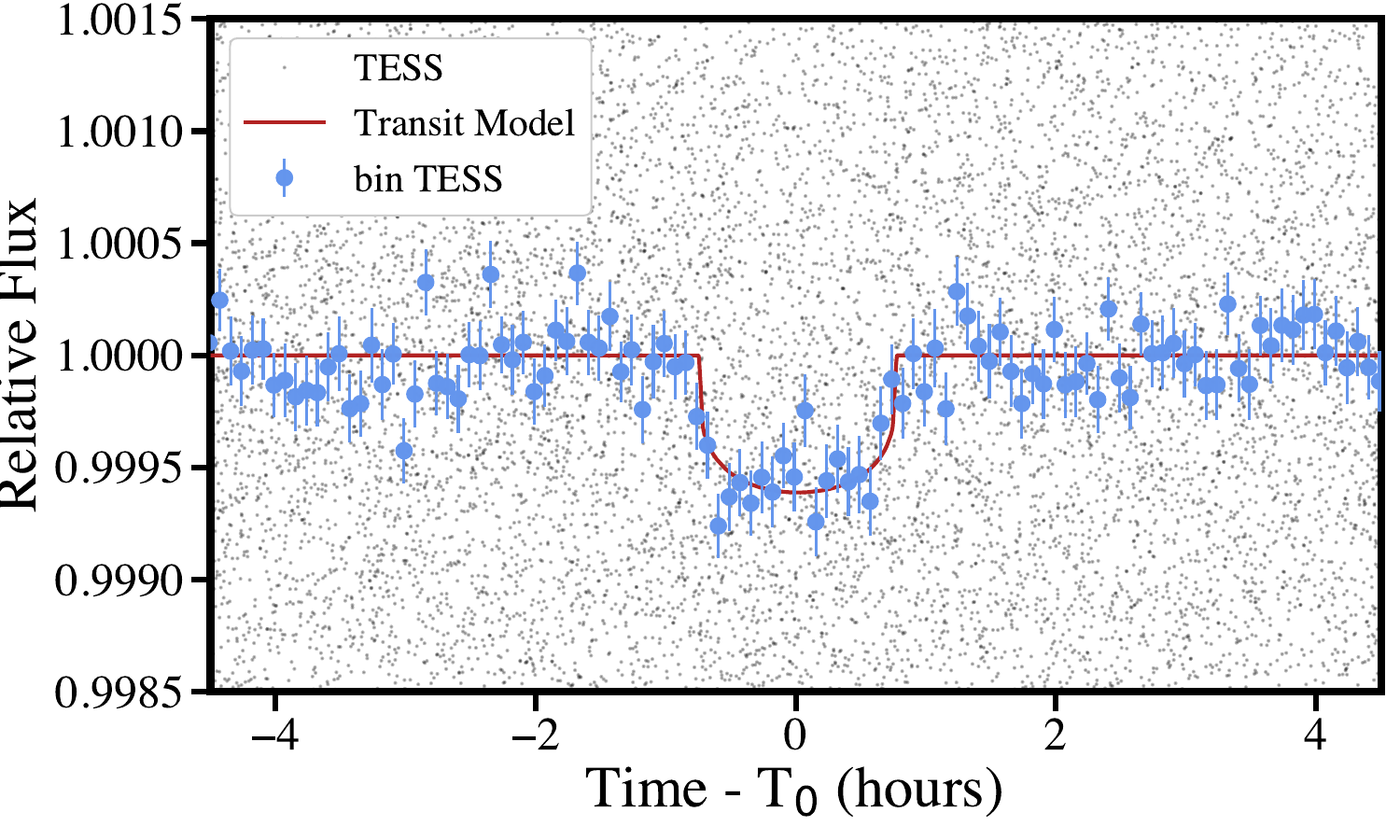}
\includegraphics[width=0.49\textwidth]{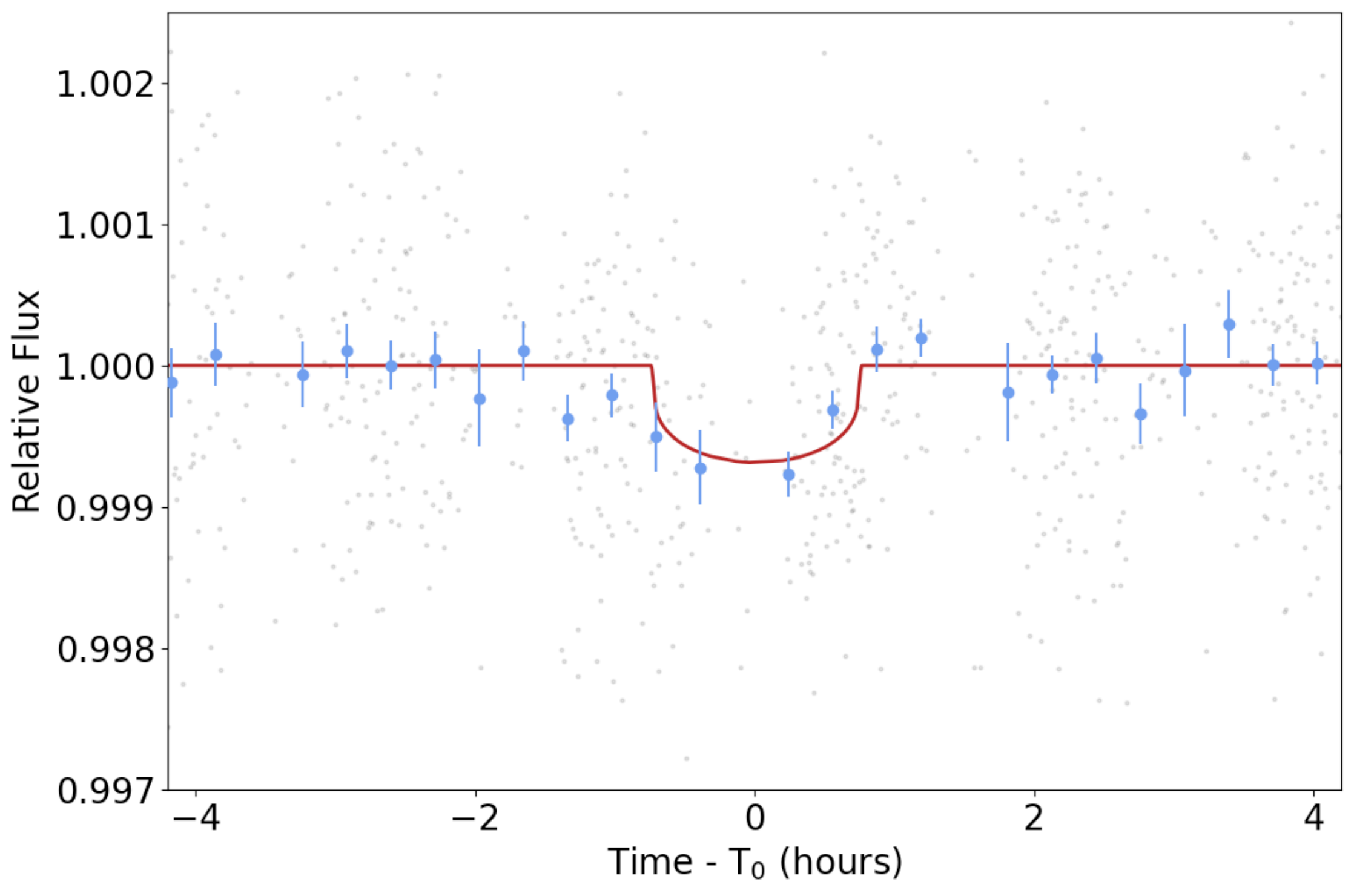}
\includegraphics[width=0.49\textwidth]{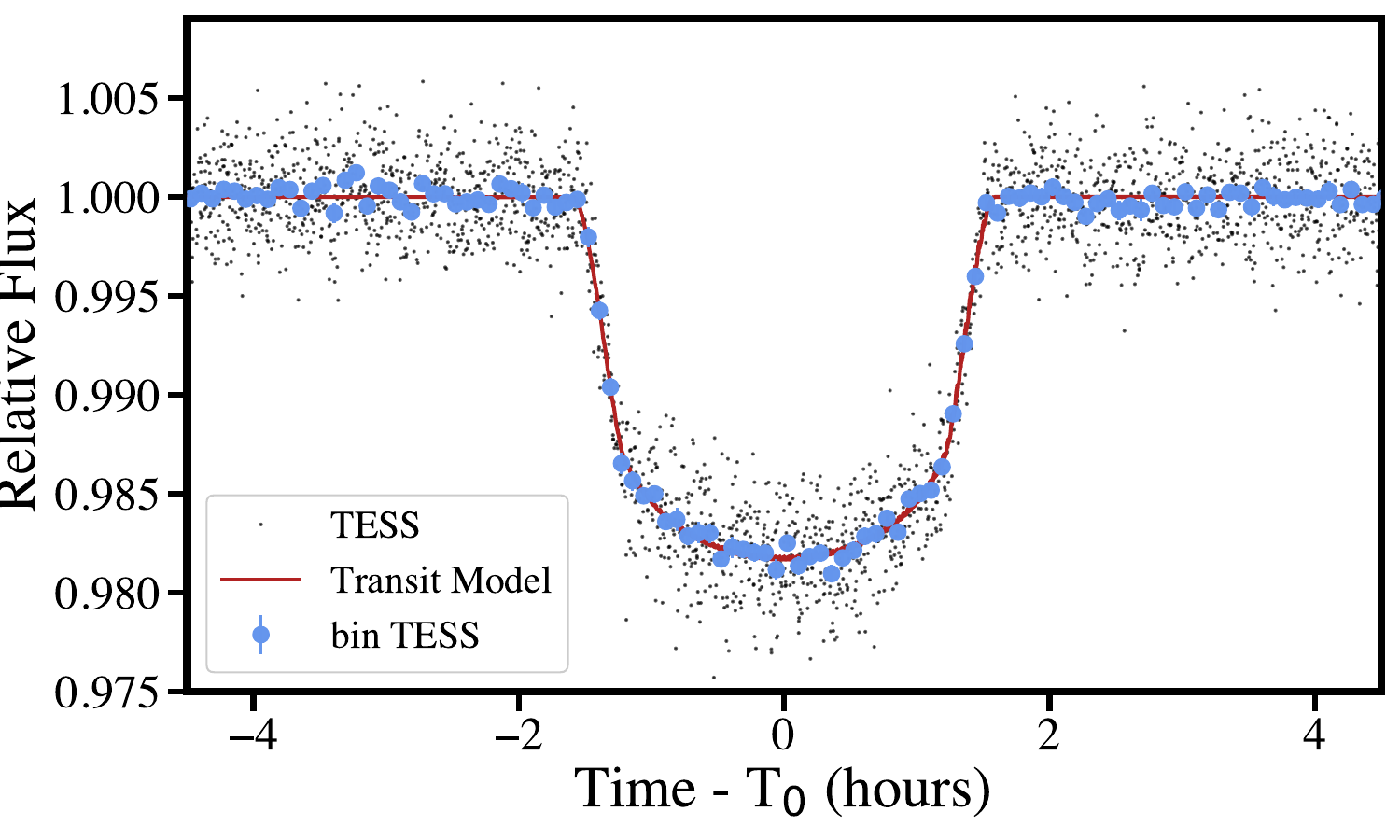}
 \caption{WASP-132 TESS and CHEOPS photometry data. Phased fit to WASP-132\,c with all three TESS sectors (top). Phased fit to WASP-132\,c with CHEOPS (middle). Phased fit to WASP-132\,b with all three TESS sectors (bottom). The transit fit from \texttt{juliet} is described in Sect. \ref{sec:juliet}. 5-minute TESS bins and 20-minute CHEOPS bins are shown in blue for display purposes.}
  \label{fig:juliet_transit_phase}
\end{figure}

\begin{figure}
  \centering
\includegraphics[width=0.49\textwidth]{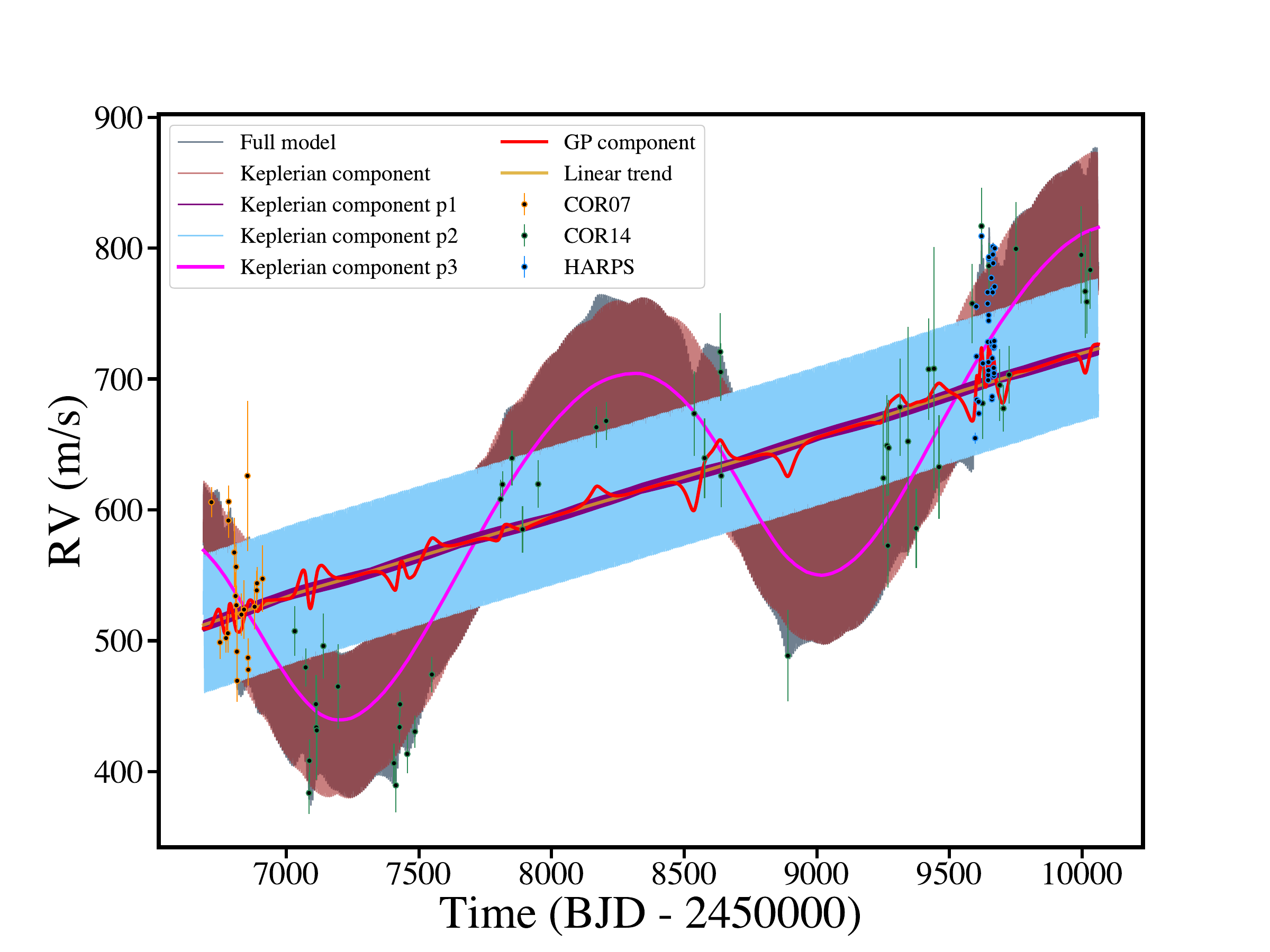}
\includegraphics[width=0.39\textwidth]{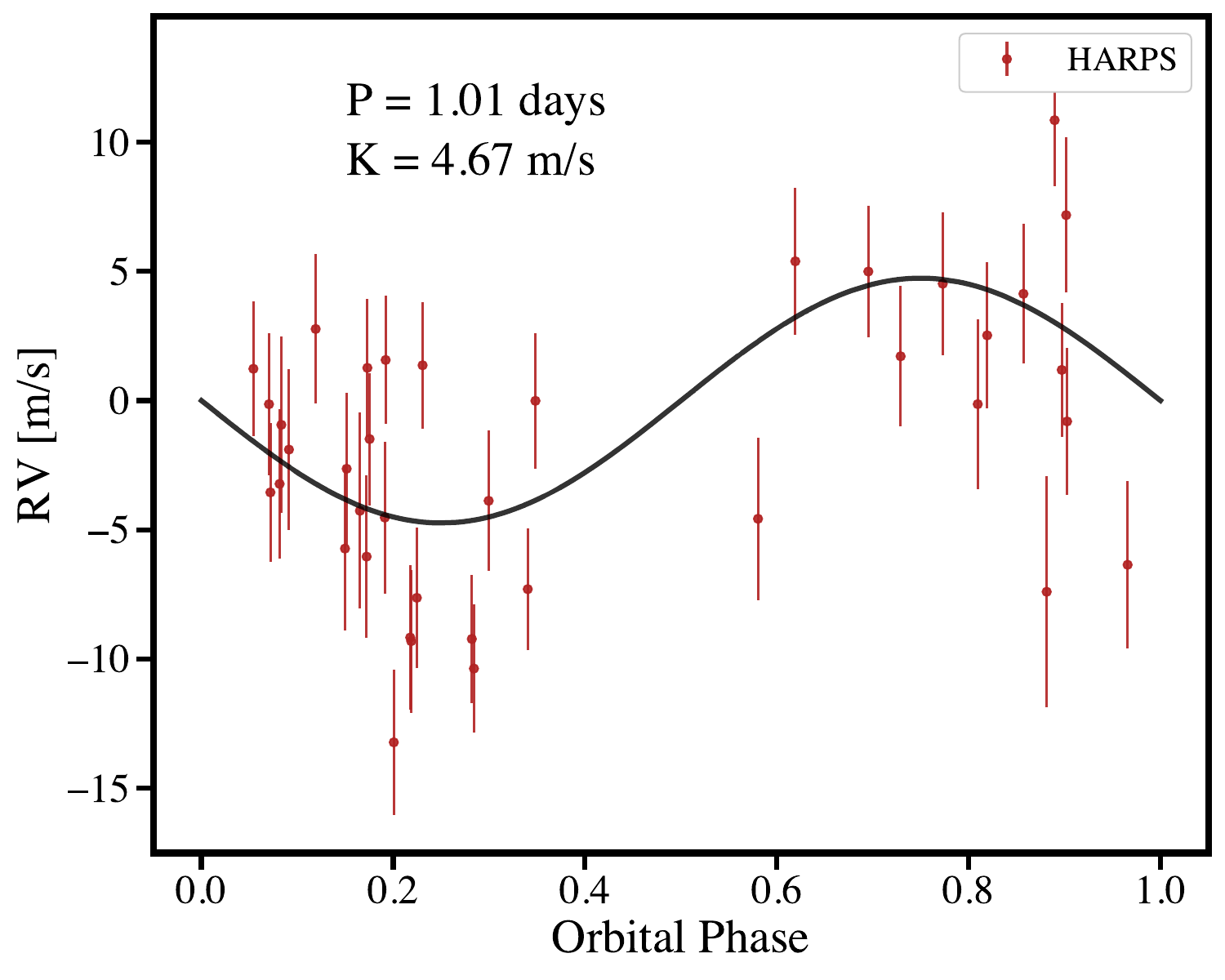}
\includegraphics[width=0.39\textwidth]{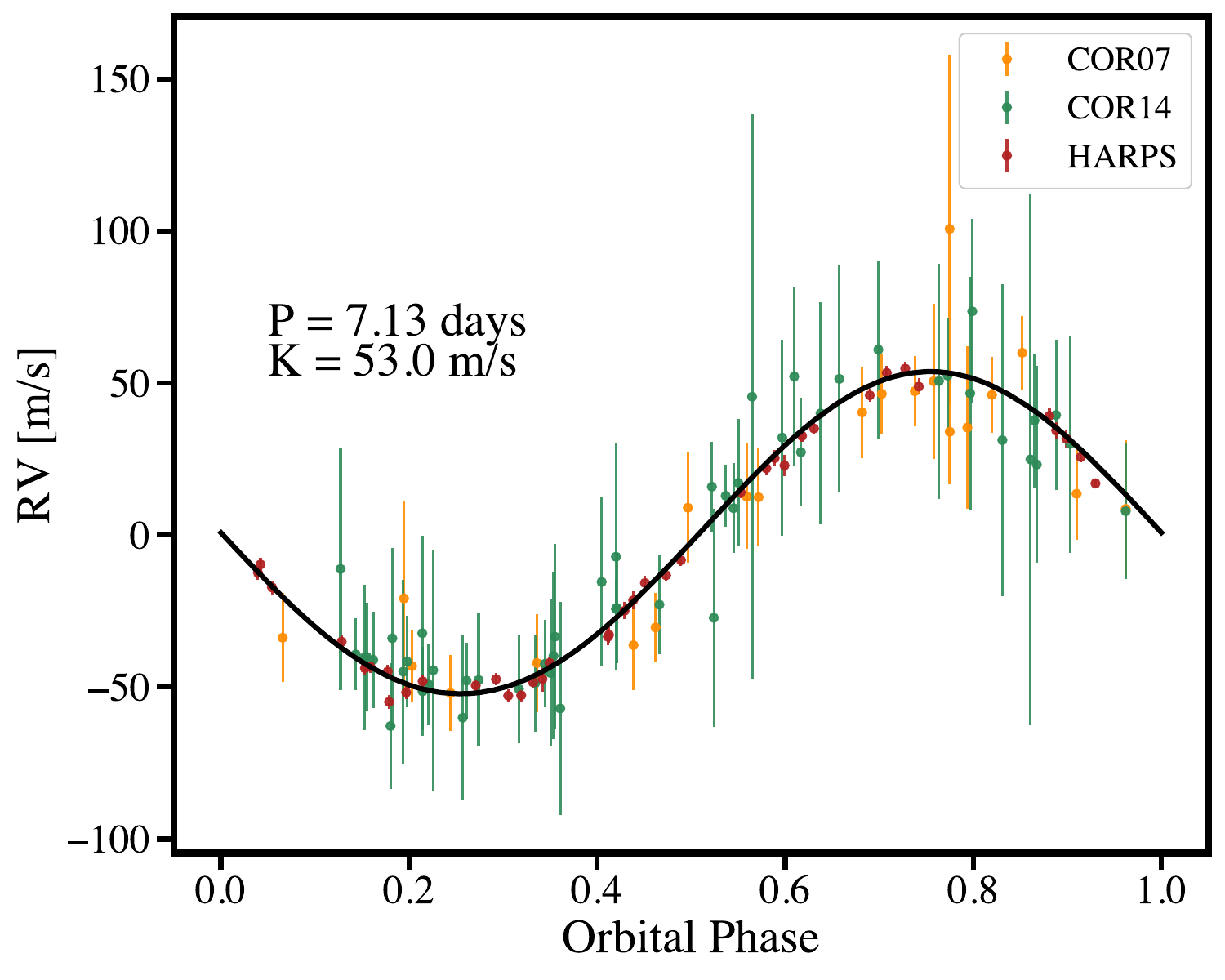}
\includegraphics[width=0.39\textwidth]{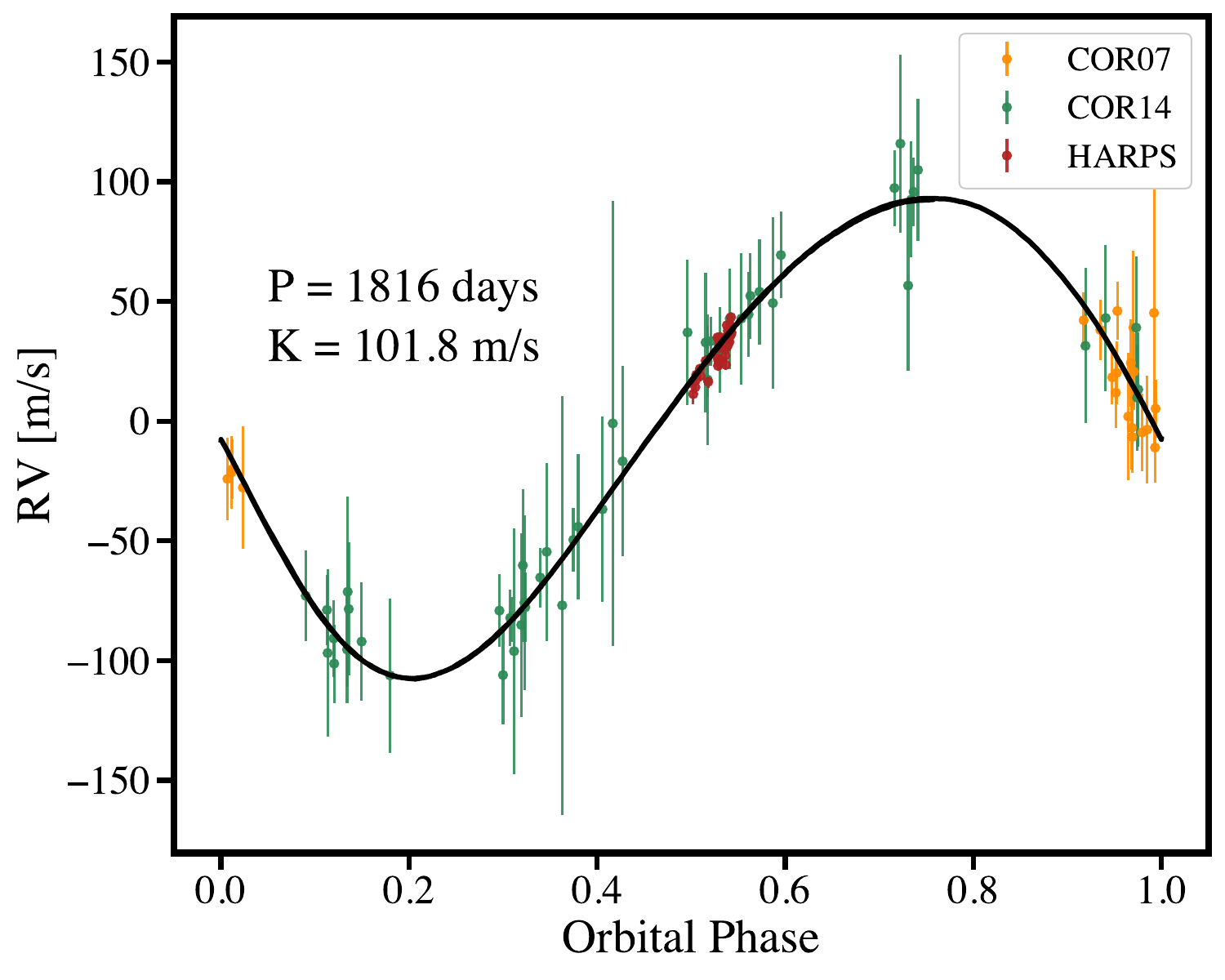}
 \caption{RV fit from global \texttt{juliet} model to WASP-132. The top plot shows the full RV set with the different model components. The second from the top shows the phased RV data and \texttt{juliet} fit to planet c with only the HARPS data plotted, while the bottom two plots show planets b and d, respectively, with both HARPS and CORALIE data.}
  \label{fig:juliet_rvs}
\end{figure}

\subsection{RV signal and activity analysis} \label{sec:rv_sig_act}

Here we further investigate the nature of the signals in the CORALIE and HARPS RVs. We employ a form of the pre-whitening procedure \citep[e.g.,][]{Queloz2009,Hatzes2013,Perger2017} based on a Fourier transform (FT) frequency analysis. In this procedure, we subtract RV signals, removing their corresponding FT peaks from the data, and then analyze the resulting generalized Lomb-Scargle (GLS) periodograms in the frequency domain. We also created spectral window functions \citep{Roberts1987,Dawson2010}, the Fourier transform of the observation times, for the CORALIE and HARPS RVs. The window function can help identify aliases in time series data as detailed by \citet{Dawson2010}.

In the top panel of Fig. 3 of the Supplementary Material published on \href{https://doi.org/10.5281/zenodo.14271139}{Zenodo}, we present a GLS periodogram in the frequency domain for the window function of the CORALIE RVs. The second panel displays the GLS periodogram of the CORALIE RVs. We account for the RV offsets between COR07 and COR14. The third through sixth panels display the GLS periodograms of the CORALIE RVs subtracted by the long-term linear trend; the linear trend and the Doppler signal of planet b; the linear trend and the Doppler signal of planet d; and the linear trend and the Doppler signals of planet d and b; respectively. From this analysis we find a strong signal at planet d's corresponding frequency after removing the linear trend and planet b's Doppler signal. We also see the FT peaks around 1 day$^{-1}$ are removed after subtracting the linear trend and planet d.


In addition, we examined GLS periodograms of the CORALIE activity indicators. The top panel of Fig. 4 of the Supplementary Material published on \href{https://doi.org/10.5281/zenodo.14271139}{Zenodo} again displays the GLS periodogram for the window function of the CORALIE data. Panels two through seven display display GLS periodograms of the CCF FWHM, CCF BIS, CCF Contrast, and the chromospheric indices measured in H$\alpha$, Na, and Ca~\textsc{ii}~H\&K lines lines, respectively. The offsets between the COR07 and COR13 activity indicators are negligible. From the CORALIE activity indicators we do not examine any significant peaks around planet d's frequency that would suggest it originated from stellar activity. We also do not see any long-term trend originating from a magnetic cycle that could explain the linear RV trend. 

We performed similar procedures with the HARPS data as displayed in Figs. 5 and 6 of the Supplementary Material published on \href{https://doi.org/10.5281/zenodo.14271139}{Zenodo}. Figure 5 of the Supplementary Material published on \href{https://doi.org/10.5281/zenodo.14271139}{Zenodo} displays GLS periodograms in the frequency domain of the HARPS window function, HARPS RVs, and the RVs with various signals subtracted including the linear trend and Doppler signals of planets d, b, and c. For the activity signal we fit the phase and amplitude of sinusoidal signals at frequencies of the approximate stellar rotation period of 31.4 days and its first two harmonics of 15.7 and 10.5 days. Although the signal is not significant, we see a small peak around planet c's frequency after subtracting the other signals and the activity signals. The GLS periodograms in the frequency domain of the HARPS activity indicators are displayed in Fig. 6 of the Supplementary Material published on \href{https://doi.org/10.5281/zenodo.14271139}{Zenodo} which includes the Mt Wilson S index S$_{MW}$. We detect significant signals around the stellar rotation frequency of $\sim$1/31.4 days in the CCF FWHM, CCF Contrast, H$\alpha$, Na, Ca~\textsc{ii}, and S$_{MW}$ GLS periodograms.

\subsection{Joint photometry and RV analysis} \label{sec:juliet}

We modeled both the photometry and RV data with \texttt{juliet} \citep{Espinoza2019}, which uses Bayesian inference to model a set number of planetary signals using {\sc \tt batman}\xspace \citep{Kreidberg2015} to model the planetary transits and \texttt{radvel} \citep{Fulton2018} to model the RVs. Of note, although \texttt{juliet} cannot handle multidimensional Gaussian processes (GPs) for RV data, for light curves \texttt{juliet} can model stellar activity as well as instrumental systematics with GPs \citep[e.g.,][]{Gibson2014} or simpler parametric functions, which we do not find available in \texttt{pyaneti}. For the transit model, \texttt{juliet} performs an efficient parameterization by fitting for the parameters $r_{1}$ and $r_{2}$ to ensure uniform exploration of the $p$ (planet-to-star ratio; $R_p/R_{\star}$) and $b$ (impact parameter) parameter space. \citet{Espinoza2018} details the $r_{1}$ and $r_{2}$ parameterization and the transformation to the $p$ and $b$ plane. We used the nested sampling method \texttt{dynesty} \citep{Speagle2019} implemented in \texttt{juliet} with 1500 live points and we ran the fit until the estimated uncertainty on the log-evidence was smaller than 0.1. 


We added a white-noise jitter term in quadrature to the error bars of both the photometry and RV data to account for underestimated uncertainties and additional noise that was not captured by the model. To account for possible residual systematics and activity affecting the transit fit of TESS Sectors 11, 38, and 65, we fit a Gaussian process (GP) using a Mat\'ern-3/2 kernel via {\sc \tt celerite}\xspace \citep{Foreman-Mackey2017} within the \texttt{juliet} framework. In addition to the stellar activity fitting presented in Sect. \ref{sec:initrv}, a GP can account for correlated noise of various origins and propagates the uncertainty \citep[e.g.,][]{Gibson2012,Gibson2014}. Additionally, properly fitting out of transit data will more accurately set the baseline for in transit data. We display the GP fit to the full TESS light curves in Fig. 1 of the Supplementary Material published on \href{https://doi.org/10.5281/zenodo.14271139}{Zenodo}. We also fit a Gaussian process (GP) using a Mat\'ern-3/2 kernel for the RV data and put uniform priors on the time-scale based on our initial RV analysis in Sect. \ref{sec:fullrv}. We fit the CHEOPS data by detrending with the background flux and the roll angle $\phi$ by using cos($\phi$), cos($2\phi$), sin($\phi$), and sin(2$\phi$). We display our full fit to the CHEOPS data in Fig. 2 of the Supplementary Material published on \href{https://doi.org/10.5281/zenodo.14271139}{Zenodo}.


\begin{table*}
\centering
\begin{minipage}{16cm}
\small
\caption{WASP-132 planet parameters from \texttt{juliet}: median and 68\% confidence interval.}
\begin{tabular}{lllc}\hline 

\hline\hline

\noalign{\smallskip}
Parameter       & &    Prior distribution\raggedright* & Value     \\
\hline

\underline{Planet c} & & & \\ 
~~~~$P$ \dotfill & Period (days) \dotfill & $\mathcal{N}(1.0115,0.001)$\dotfill & $1.01153624^{+0.00000093}_{-0.00000086}$ \\
~~~~$T_0$ \dotfill & Time of transit center (BJD$_{\text{TDB}}$) \dotfill & $\mathcal{N}(2458597.576,0.001)$\dotfill & $2458597.57584^{+0.00068}_{-0.00067}$ \\
~~~~$e$\dotfill & Eccentricity of the orbit \dotfill & $fixed$\dotfill & 0 \\
~~~~$K$\dotfill & Radial velocity semi-amplitude (\ms) \dotfill & $\mathcal{N}(4.60,1.37)$\dotfill & $4.67^{+1.37}_{-1.37}$$\dagger$ \\
~~~~$\omega$\dotfill & Argument of periastron (deg) \dotfill & $\mathcal{U}(0,360)$\dotfill & $188^{+106}_{-113}$ \\
~~~~$r_1$\dotfill & Parametrization for p and b \dotfill & $\mathcal{U}(0,1)$\dotfill & $0.405^{+0.041}_{-0.041}$ \\
~~~~$r_2$\dotfill & Parametrization for p and b \dotfill & $\mathcal{U}(0,1)$\dotfill & $0.02230^{+0.00060}_{-0.00065}$ \\
~~~~$i$\dotfill & Inclination (deg) \dotfill & \dotfill & $88.82^{+0.67}_{-0.69}$ \\
~~~~$p=R_p/R_{\star}$\dotfill & Planet-to-star radius ratio \dotfill & \dotfill & $0.02230^{+0.00060}_{-0.00065}$ \\
~~~~$b$\dotfill &Impact parameter of the orbit \dotfill & \dotfill & $0.107^{+0.062}_{-0.061}$ \\
~~~~$a$\dotfill & Semi-major axis (AU)  \dotfill & \dotfill & $0.01833^{+0.00079}_{-0.00079}$ \\
~~~~$M_p$\dotfill & Planetary mass (M$_{\oplus}$)  \dotfill & \dotfill & $6.26^{+1.84}_{-1.83}$ \\
~~~~$R_p$\dotfill & Planetary radius ($R_{\oplus}$)  \dotfill & \dotfill & $1.841^{+0.094}_{-0.093}$ \\
~~~~$\rho _p$\dotfill & Planetary density (g\,cm$^{-3}$)  \dotfill & \dotfill & $5.47^{+1.96}_{-1.71}$ \\
~~~~$S$\dotfill & Insolation ($S_{\oplus}$)  \dotfill & \dotfill & $791^{+72}_{-64}$ \\
~~~~$T_{eq}$\dotfill & Equilibrium Temperature ($K$) \dotfill & \dotfill & $1329^{+123}_{-173}$ \\
~~~~$T_{\rm{dur}}$\dotfill & Total Transit Duration (hours)  \dotfill & \dotfill & $1.491^{+0.016}_{-0.016}$ \\
\underline{Planet b} & & & \\ 
~~~~$P$ \dotfill & Period (days) \dotfill & $\mathcal{U}$(7.1335$\pm$0.1)\dotfill & $7.1335164^{+0.0000019}_{-0.0000019}$ \\
~~~~$T_0$ \dotfill & Time of transit center (BJD$_{\text{TDB}}$) \dotfill & $\mathcal{U}$(2459337.6$\pm$0.2)\dotfill & $2459337.60808^{+0.00014}_{-0.00016}$ \\
~~~~$e$\dotfill & Eccentricity of the orbit \dotfill & $\mathcal{B}(0.867,3.03)$\dotfill & $0.0163^{+0.0067}_{-0.0069}$ \\
~~~~$K$\dotfill & Radial velocity semi-amplitude (\ms) \dotfill & $\mathcal{U}(0,1000)$\dotfill & $52.99^{+0.75}_{-0.74}$ \\
~~~~$\omega$\dotfill & Argument of periastron (deg) \dotfill & $\mathcal{U}(0,360)$\dotfill & $318^{+31}_{-298}$ \\
~~~~$r_1$\dotfill & Parametrization for p and b \dotfill & $\mathcal{U}(0,1)$\dotfill & $0.4535^{+0.0250}_{-0.0281}$ \\
~~~~$r_2$\dotfill & Parametrization for p and b \dotfill & $\mathcal{U}(0,1)$\dotfill & $0.12219^{+0.00060}_{-0.00052}$ \\
~~~~$i$\dotfill & Inclination (deg) \dotfill & \dotfill & $89.46^{+0.13}_{-0.11}$ \\
~~~~$p=R_p/R_{\star}$\dotfill & Planet-to-star radius ratio \dotfill & \dotfill & $0.12219^{+0.00050}_{-0.00052}$ \\
~~~~$b$\dotfill &Impact parameter of the orbit \dotfill & \dotfill & $0.180^{+0.037}_{-0.042}$ \\
~~~~$a$\dotfill & Semi-major axis (AU)  \dotfill & \dotfill & $0.0674^{+0.0029}_{-0.0029}$ \\
~~~~$M_p$\dotfill & Planetary mass ($M_{\rm{Jup}}$)  \dotfill & \dotfill & $0.428^{+0.015}_{-0.015}$ \\
~~~~$R_p$\dotfill & Planetary radius ($R_{\rm{Jup}}$)  \dotfill & \dotfill & $0.901^{+0.038}_{-0.038}$ \\
~~~~$\rho _p$\dotfill & Planetary density (g\,cm$^{-3}$)  \dotfill & \dotfill & $0.73^{+0.10}_{-0.09}$ \\
~~~~$S$\dotfill & Insolation ($S_{\oplus}$)  \dotfill & \dotfill & $58.5^{+5.3}_{-4.7}$ \\
~~~~$T_{eq}$\dotfill & Equilibrium Temperature ($K$) \dotfill & \dotfill & $682^{+75}_{-114}$ \\
~~~~$T_{\rm{dur}}$\dotfill & Total Transit Duration (hours)  \dotfill & \dotfill & $11.04^{+0.18}_{-0.18}$ \\
\underline{Planet d} & & & \\ 
~~~~$P$ \dotfill & Period (days) \dotfill & $\mathcal{N}(1812,44)$\dotfill & $1816.6^{+44.4}_{-44.4}$$\dagger$ \\
~~~~$T_0$ \dotfill & Time of inferior conjunction (BJD$_{\text{TDB}}$) \dotfill & $\mathcal{N}(2455059.5,95)$\dotfill & $2455051.5^{+95.0}_{-95.0}$$\dagger$ \\
~~~~$e$\dotfill & Eccentricity of the orbit \dotfill & $\mathcal{N}(0.1,0.08)$\dotfill & $0.120^{+0.078}_{-0.078}$$\dagger$ \\
~~~~$K$\dotfill & Radial velocity semi-amplitude (\ms) \dotfill & $\mathcal{N}(104,9.6)$\dotfill & $101.81^{+9.60}_{-9.60}$$\dagger$ \\
~~~~$\omega$\dotfill & Argument of periastron (deg) \dotfill & $\mathcal{N}(135,40)$\dotfill & $131^{+40}_{-40}$$\dagger$ \\
~~~~$a$\dotfill & Semi-major axis (AU)  \dotfill & \dotfill & $2.71^{+0.12}_{-0.12}$ \\
~~~~$M_p$ sin$i$\dotfill & Planetary mass $\times$ sine inclination ($M_{\rm{Jup}}$)  \dotfill & \dotfill & $5.16^{+0.52}_{-0.52}$ \\
~~~~$S$\dotfill & Insolation ($S_{\oplus}$)  \dotfill & \dotfill & $0.0363^{+0.0034}_{-0.0030}$ \\
~~~~$T_{eq}$\dotfill & Equilibrium Temperature ($K$) \dotfill & \dotfill & $107^{+11}_{-18}$ \\

\hline
\end{tabular}
\label{tab:juliet_physical}
{\\ \raggedright*$\mathcal{U}(a,b)$ indicates a uniform distribution between $a$ and $b$; $\mathcal{J}(a,b)$ a Jeffrey or log-uniform distribution between $a$ and $b$; $\mathcal{N}(a,b)$ a normal distribution with mean $a$ and standard deviation $b$; and $\mathcal{B}(a,b)$ a Beta prior as detailed in \citet{Kipping2014}. Parameters with no prior distribution were derived. We sample from a normal distribution for the stellar mass, stellar radius, and stellar temperature, that are based on the results from Sect. \ref{sec:star} to derive parameters. \\${\dagger}$Uncertainties account for the priors set from the initial RV fit. \\
}
\end{minipage}
\end{table*}

\begin{table*}
\centering
\begin{minipage}{16cm}
\small
\caption{WASP-132 system and instrumental parameters from \texttt{juliet}: median and 68\% confidence interval.}
\begin{tabular}{lllc}\hline 
\hline\hline

\noalign{\smallskip}
Parameter       & &    Prior distribution\raggedright* & Value     \\
\hline

~~~~$\rho _{*}$\dotfill & Stellar density (g\,cm$^{-3}$)  \dotfill & $\mathcal{N}(2.56,0.21)$\dotfill & $2.602^{+0.067}_{-0.070}$ \\
~~~~$q_{\text{1,TESS}}$\dotfill & Quadratic limb-darkening parametrization\dotfill & $\mathcal{N}(0.429,0.017)$\dotfill & $0.4228^{+0.0155}_{-0.0152}$ \\
~~~~$q_{\text{2,TESS}}$\dotfill & Quadratic limb-darkening parametrization\dotfill & $\mathcal{N}(0.370,0.025)$\dotfill & $0.3597^{+0.0177}_{-0.0182}$ \\
~~~~$q_{\text{1,CHEOPS}}$\dotfill & Quadratic limb-darkening parametrization\dotfill & $\mathcal{N}(0.529,0.019)$\dotfill & $0.5245^{+0.0153}_{-0.0153}$ \\
~~~~$q_{\text{2,CHEOPS}}$\dotfill & Quadratic limb-darkening parametrization\dotfill & $\mathcal{N}(0.428,0.022)$\dotfill & $0.4272^{+0.0202}_{-0.0193}$ \\
~~~~$m_{\text{flux,TESS11}}$\dotfill & Offset (relative flux)\dotfill & $\mathcal{N}(0,0.01)$\dotfill & $-0.0076^{+0.0059}_{-0.0060}$ \\
~~~~$m_{\text{flux,TESS38}}$\dotfill & Offset (relative flux)\dotfill & $\mathcal{N}(0,0.01)$\dotfill &  $0.0082^{+0.0073}_{-0.0060}$ \\
~~~~$m_{\text{flux,TESS65}}$\dotfill & Offset (relative flux)\dotfill & $\mathcal{N}(0,0.01)$\dotfill &  $-0.0020^{+0.0064}_{-0.0064}$ \\
~~~~$m_{\text{flux,CHEOPS}}$\dotfill & Offset (relative flux)\dotfill & $\mathcal{N}(0,0.01)$\dotfill &  $-0.00013^{+0.00012}_{-0.00012}$ \\
~~~~$\sigma_{\text{TESS11}}$\dotfill  & Jitter (ppm)\dotfill & $\mathcal{J}(10^{-5},100.0)$\dotfill & $0.086^{+5.820}_{-0.085}$ \\
~~~~$\sigma_{\text{TESS38}}$\dotfill  & Jitter (ppm)\dotfill & $\mathcal{J}(10^{-5},100.0)$\dotfill & $0.034^{+3.102}_{-0.033}$ \\
~~~~$\sigma_{\text{TESS65}}$\dotfill  & Jitter (ppm)\dotfill & $\mathcal{J}(10^{-5},100.0)$\dotfill & $0.106^{+6.306}_{-0.106}$ \\
~~~~$\sigma_{\text{CHEOPS}}$\dotfill  & Jitter (ppm)\dotfill & $\mathcal{J}(10^{-5},100.0)$\dotfill & $0.0029^{+0.1605}_{-0.0028}$ \\
~~~~$\sigma_{\text{GP,TESS11}}$\dotfill & GP amplitude (relative flux)\dotfill & $\mathcal{J}(10^{-6},1)$\dotfill &  $0.0155^{+0.0053}_{-0.0043}$ \\
~~~~$\sigma_{\text{GP,TESS38}}$\dotfill & GP amplitude (relative flux)\dotfill & $\mathcal{J}(10^{-6},1)$\dotfill &  $0.0240^{+0.0064}_{-0.0056}$ \\
~~~~$\sigma_{\text{GP,TESS65}}$\dotfill & GP amplitude (relative flux)\dotfill & $\mathcal{J}(10^{-6},1)$\dotfill &  $0.0895^{+0.0265}_{-0.0236}$ \\
~~~~$\rho_{\text{GP,TESS11}}$\dotfill & Photometry GP time-scale (days)\dotfill & $\mathcal{N}(31.3,5.0)$\dotfill &  $30.16^{+3.96}_{-4.72}$ \\
~~~~$\rho_{\text{GP,TESS38}}$\dotfill & Photometry GP time-scale (days)\dotfill & $\mathcal{N}(31.3,5.0)$\dotfill &  $30.41^{+4.12}_{-4.31}$ \\
~~~~$\rho_{\text{GP,TESS65}}$\dotfill & Photometry GP time-scale (days)\dotfill & $\mathcal{N}(31.3,5.0)$\dotfill &  $29.21^{+4.03}_{-5.40}$ \\
~~~~$\theta_{0\text{CHEOPS}}$\dotfill  & Detrending Parameter \dotfill & $\mathcal{U}(-0.1,0.1)$\dotfill & $0.000566^{+0.000084}_{-0.000077}$ \\
~~~~$\theta_{1\text{CHEOPS}}$\dotfill  & Detrending Parameter \dotfill & $\mathcal{U}(-0.1,0.1)$\dotfill & $-0.000512^{+0.000069}_{-0.000068}$ \\
~~~~$\theta_{2\text{CHEOPS}}$\dotfill  & Detrending Parameter \dotfill & $\mathcal{U}(-0.1,0.1)$\dotfill & $0.000936^{+0.000104}_{-0.000099}$ \\
~~~~$\theta_{3\text{CHEOPS}}$\dotfill  & Detrending Parameter \dotfill & $\mathcal{U}(-0.1,0.1)$\dotfill & $0.000240^{+0.000046}_{-0.000046}$ \\
~~~~$\theta_{4\text{CHEOPS}}$\dotfill  & Detrending Parameter \dotfill & $\mathcal{U}(-0.1,0.1)$\dotfill & $0.001940^{+0.000671}_{-0.000688}$ \\
~~~~$\sigma_{\text{GP,RV}}$\dotfill & RV GP amplitude ($\ms$)\dotfill & $\mathcal{U}(0.01,100)$\dotfill &  $42.48^{+6.43}_{-5.91}$ \\
~~~~$\rho_{\text{GP,RV}}$\dotfill & RV GP time-scale (days)\dotfill & $\mathcal{U}(29.3,33.3)$\dotfill &  $30.03^{+0.81}_{-0.54}$ \\
~~~~$\mu_{\text{COR07}}$\dotfill & Systemic RV offset (\kms)\dotfill & $\mathcal{U}(-100,100)$\dotfill & $30.505^{+0.333}_{-0.267}$ \\
~~~~$\sigma_{\text{COR07}}$\dotfill  & Jitter (\ms)\dotfill &  $\mathcal{U}(0,2)$\dotfill & $0.968^{+0.580}_{-0.561}$ \\
~~~~$\mu_{\text{COR14}}$\dotfill & Systemic RV offset (\kms)\dotfill & $\mathcal{U}(-100,100)$\dotfill & $30.512^{+0.326}_{-0.259}$ \\
~~~~$\sigma_{\text{COR14}}$\dotfill  & Jitter (\ms)\dotfill &  $\mathcal{U}(0,2)$\dotfill & $0.94^{+0.59}_{-0.54}$ \\
~~~~$\mu_{\text{HARPS}}$\dotfill & Systemic RV offset (\kms)\dotfill &  $\mathcal{U}(-100,100)$\dotfill & $30.551^{+0.327}_{-0.257}$ \\
~~~~$\sigma_{\text{HARPS}}$\dotfill  & Jitter (\ms)\dotfill &  $\mathcal{U}(0,2)$\dotfill & $1.63^{+0.25}_{-0.40}$ \\
~~~~A$_{RV}$\dotfill  & slope of linear long-term RV trend (\ms day$^{-1}$) &  $\mathcal{N}(0.0645,0.01)$\dotfill & $0.0648^{+0.0063}_{-0.0063}$ \\
~~~~B$_{RV}$\dotfill  & intercept of linear long-term RV trend (\ms) &  $\mathcal{U}(-1000,1000)$\dotfill & $611.9^{+260.6}_{-326.9}$ \\

\hline
\end{tabular}
\label{tab:juliet_instrument}
{\\ \raggedright*Prior symbols are the same as those described in Table \ref{tab:juliet_physical}.
}
\end{minipage}
\end{table*}

For limb darkening, we derived quadratic coefficients and their uncertainties for different photometric filters using the \texttt{LDCU}$\footnote{\url{https://github.com/delinea/LDCU}}$ routine \citep{Deline2022} and used them to set Gaussian priors on the limb-darkening parameters. We used uniform priors for the period and transit center times. We fixed the eccentricity for the 1-day period planet and added a Beta prior for transiting planets for the hot Jupiter, as described in \citet{Kipping2014}. We used the $\rho_*$ (stellar density) as a parameter instead of the scaled semi-major axis ($\it{a}$/R$_{\star}$). The normal prior on stellar density is informed by the stellar analysis in Sect. \ref{sec:star}.

For our final RV and photometry joint fit, we constrained the RV semi-amplitude $K$ of planet c as well as the period, $T_0$, eccentricity, RV $K$, and $\omega$ of planet d using values from our initial \texttt{pyaneti} RV analysis presented in Sect. \ref{sec:initrv}. We tested several times putting fewer constraints on these parameters within the \texttt{juliet} global model but found that the model would get stuck with solutions that had lower likelihoods than our final solution. We therefore used priors from our initial \texttt{pyaneti} analysis to guide the \texttt{juliet} global model to find the most likely solution. We account for these priors in our uncertainties of the modeled and derived \texttt{juliet} parameters.

Tables \ref{tab:juliet_physical} and \ref{tab:juliet_instrument} display all of the modeled parameters as well as their input priors for our joint RV and transit model. Figure \ref{fig:juliet_transit_phase} displays the final model fits to the phased photometry data for both WASP-132\,c and WASP-132\,b. Figure \ref{fig:juliet_rvs} displays the final RV fits with the joint RV and photometry model. We find the planet masses to be consistent with our RV-only analysis and therefore present the joint fit as our final parameters.

In addition to all of the modeled parameters, Table \ref{tab:juliet_physical} also displays derived planet parameters including the inclination, impact parameter, semi-major axis, and radius. We calculated the insolation using the equation:
\begin{equation}
    S [S_{\oplus}] = L_{*} [L_{\odot}] \left(a [\rm{AU}] \right)^{-2}.
\end{equation}
We calculated the equilibrium temperature assuming a Bond albedo of $A$\,=\,0.343 (the same as Jupiter's) for the hot Jupiter and $A$\,=\,0.3 (the same as Earth's) for the super-Earth and the semi-major axis distance $a$ using the equation:
\begin{equation}
T_{eq} = T_{\rm{eff}} (1-A)^{1/4}\sqrt{\frac{R_{\star}}{2a}}.
\end{equation}
We set upper and lower uncertainties for the equilibrium temperature by assuming Bond albedos of $A$\,=\,0 and $A$\,=\,0.686 (double that of Jupiter), respectively for the hot and cold Jupiters and $A$\,=\,0 and $A$\,=\,0.6 (double that of Earth), respectively for the super-Earth.

\section{Discussion} \label{sec:disc}

\subsection{Comparison to previous studies}

We find our stellar parameters in agreement with \citet{Hord2022} and other previous studies. \citet{Hord2022} found a radius of 10.05\,$\pm$\,0.28\,$R_{\oplus}$ and mass of 121.9\,$\pm$\,21.9\,$M_{\oplus}$ for WASP-132\,b, while \citet{Hellier2017} found a radius of 9.75\,$\pm$\,0.34\,$R_{\oplus}$ and mass of 130.3\,$\pm$\,9.5\,$M_{\oplus}$. Notably \citet{Hord2022} included a fit for WASP-132\,c in their RV analysis to obtain the mass of WASP-132\,b. With a measurement 136.1\,$\pm$\,4.8\,$M_{\oplus}$, we decrease by a factor of two the mass uncertainty for WASP-132\,b, and we find a similar radius of 10.10\,$\pm$\,0.43\,$R_{\oplus}$. \citet{Hord2022} found a radius of 1.85\,$\pm$\,0.10\,$R_{\oplus}$ for WASP-132\,c, which is similar to our result of 1.841\,$\pm$\,0.094\,$R_{\oplus}$. For the two inner planets we find bulk densities of $\rho_{\rm{c}}$\,=\,$5.47^{+1.96}_{-1.71}$\,g\,cm$^{-3}$ for WASP-132\,c and $\rho_{\rm{b}}$\,=\,$0.73^{+0.10}_{-0.09}$\,g\,cm$^{-3}$ for WASP-132\,b.

\subsection{Dynamical history}

With the help of previous studies such as \citealt{DawsonJohnson2018}, \citet{Wu2023} recently gave a unified framework for short-period gas giant dynamical sculpting where hot and warm Jupiters emerge as a natural outcome of postdisk dynamical sculpting of gas giants in compact multi-planet systems. They propose that after the initial starting point from the disk evolution, planet-planet interactions can excite the eccentricities of some giant planets and push them into smaller orbits. This eccentric migration continuum can be divided into three regimes of hot Jupiters, those with quiescent histories, low-eccentricity migration, and high-eccentricity migration \citep{Wu2023}. Given the presence of the small inner planet, WASP-132\,b likely formed with a quiescent history without experiencing strong interactions with other planets in the system, maintaining low eccentricity (we find e$\sim$0 for WASP-132\,b), and retaining nearby planetary companions. This scenario is more likely than low-eccentricity or high-eccentricity migration that involve interactions with other gas giants and are likely to remove nearby companions.

\begin{figure}
  \centering
\includegraphics[width=0.5\textwidth]{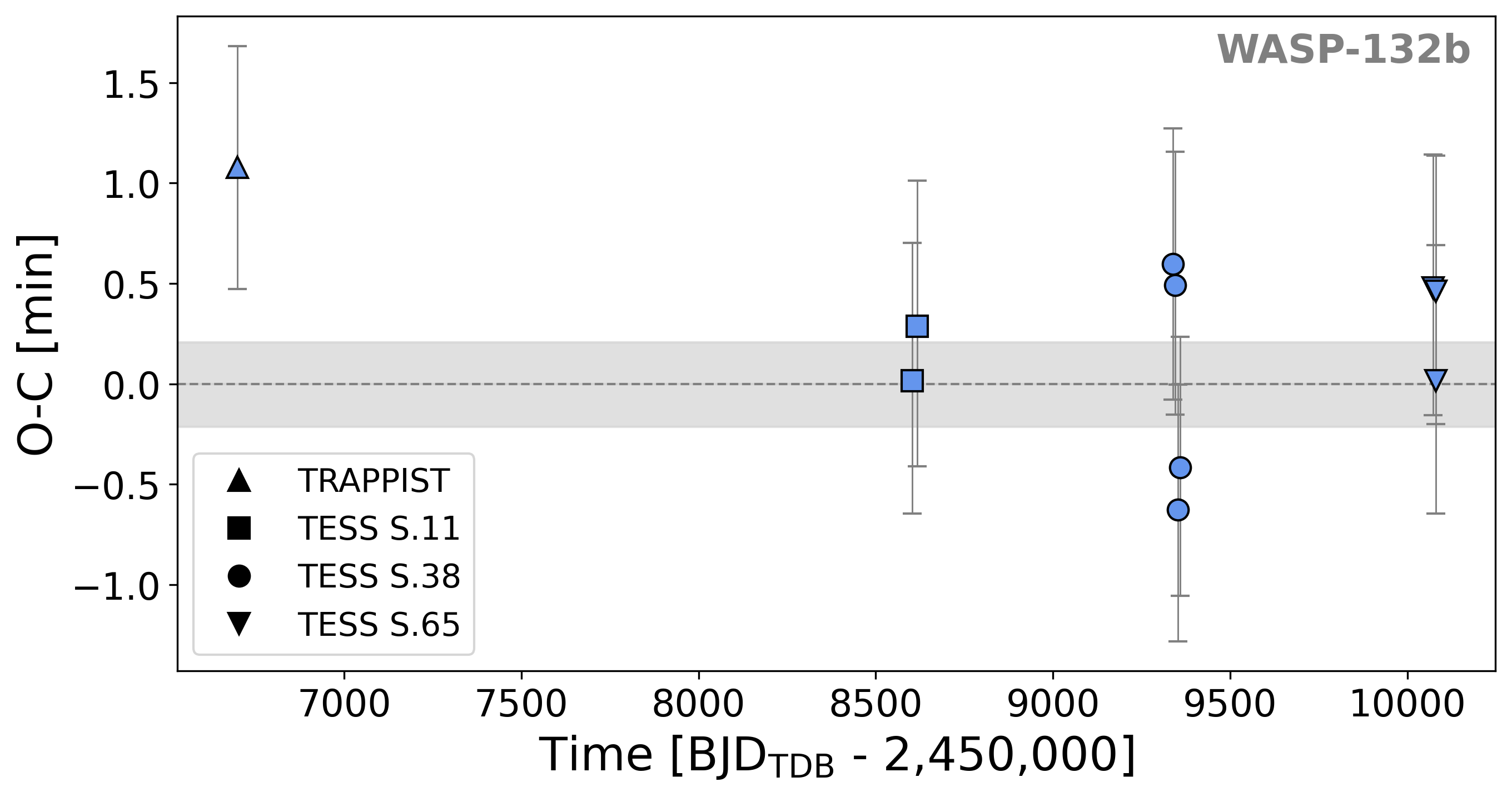}
 \caption{TTVs of WASP-132\,b determined with \texttt{juliet} on TESS data and the TRAPPIST transit from \citet{Hellier2017}. The gray area represents the 1$\sigma$ uncertainty of the $T_0$ from our global analysis.}
  \label{fig:wasp132b_ttv}
\end{figure}

Although WASP-132\,c and WASP-132\,b are not in resonance, we performed a TTV analysis of WASP-132\,b to test if any other non-transiting or non-RV detected planets were causing perturbations of the planet. We conducted an analysis of the TTVs for WASP-132\,b with \texttt{juliet} by utilizing all available photometric datasets from TESS. Instead of fitting a single period ($\it{P_{orb}}$) and time-of-transit center ($T_0$), \texttt{juliet} employed a method that seeks individual transit times. This approach involved fitting each transit independently and determining one transit time for each one, resulting in a more consistent and coherent analysis. We also included  the TRAPPIST observation of the WASP-132\,b transit made on 2014 May 5 from \citet{Hellier2017} for our TTV analysis. Figure \ref{fig:wasp132b_ttv} illustrates the results of our analysis, which compares the observed transit times with the calculated linear ephemeris derived from all the transits. Notably, no significant variations were detected within the TESS data and the TRAPPIST transit is only a $\sim$1$\sigma$ outlier.

Additionally, hot Jupiters in systems with two or more known stars could have been produced through eccentric migration triggered by secular Kozai-Lidov interactions with stellar companions \citep[e.g., HD 80606\,b;][]{Naef2001,WuMurray2003}. \citet{Hord2022} presented speckle imaging of WASP-132 using the HRCam instrument on the 4.1\,m SOAR telescope \citep{Tokovinin2018} and detected no nearby stars within 3'' of WASP-132. \citet{Hord2022} also observed WASP-132 with the LCOGT \citep{Brown2013} 1.0\,m network node at the South African Astronomical Observatory and ruled out nearby eclipsing binaries for all neighboring stars out to 50'' from WASP-132. However, in addition to the 2.7 AU outer planet, we also significantly detect a long-term trend in our CORALIE RVs. This could be due to an additional brown dwarf or low-mass star in the system. Using a basic calculation of mass from the RV span of $\sim$200\,m\,s$^{-1}$ (minimum $K$), time span of $\sim$9\,years (minimum period\,$\sim$18\,years), and assuming a circular orbit, we find a potential minimum mass of this outer companion to be $\sim$18.5\,$M_{\rm{Jup}}$. This possible additional brown dwarf or stellar companion could have influenced the migration of the two inner planets as well as the outer giant planet.

The stellar obliquity or the observed excess of spin-orbit misalignment can put important constraints on the formation pathways of hot Jupiters. If WASP-132b did migrate early on in the proto-planetary disk, measuring the spin-orbit angles of the two planets using the Rossiter-McLaughlin effect will allow the coplanarity of their orbits and their possible misalignments with the star to be assessed, which constrains their dynamical history. As noted previously, ESPRESSO spectroscopic observations of the transits of both planets were obtained and are planned for publication by a separate team.

\subsection{Astrometric signal}

Astrometry can be an ideal method to provide exact masses of companions in combination with RVs. A commonly applied approach analyses the proper motion anomaly between Hipparcos \citep{ESA1997} and Gaia \citep{GaiaCollaboration2016}, described in more details by \citet{Snellen2018} and \citet{Brandt2021}. WASP-132 was too faint to be observed by Hipparcos, which means that the comparison is not possible; however, DR3 \citep{GaiaCollaboration2023} includes an additional 12 months of observations more than DR2 \citep{GaiaCollaboration2018} and can be compared. The short-period planets are averaged into the proper motions reported by these two datasets, but the proper motion anomaly of WASP-132\,d (1817 days) should be visible, if significant. Whereas the long term trend we detected in the RVs should be seen as a constant proper motion offset and will likely not introduce any difference between DR2 and DR3. 

There is a 4-sigma difference ($\Delta$mu$_{\alpha^*}$\,=\,$-$0.30 $\pm$ 0.072 mas\,yr$^{-1}$) between the proper motion values in Right Ascension from DR3 (mu$_{\alpha^*}$\,=\,12.26\,$\pm$\,0.02 mas\,yr$^{-1}$) and DR2 (mu$_{\alpha^*}$\,=\,12.56\,$\pm$\,0.07 mas\,yr$^{-1}$). While the difference in proper motions in Declination ($\Delta$mu$_{\delta}$ = 0.04 $\pm$ 0.072 mas\,yr$^{-1}$) is not significant (DR3 mu$_{\delta}$\,=\,$-$73.17\,$\pm$\,0.02\,mas\,yr$^{-1}$ and DR2 mu$_{\delta}$\,=\,$-$73.21\,$\pm$\,0.07\,mas\,yr$^{-1}$). As the proper motion of DR3 also includes the data of DR2, the difference in proper motion could be an indication of astrometric signal but it could also be caused by improved Gaia calibration and data reduction. 

Additionally, Gaia DR3 reports a Renormalized Unit Weight Error (RUWE) of 1.23, where 1 is generally seen as a good measurement and above 1.4 as a possible indication of binarity \citep{Kervella2022}. The Gaia DR3 astrometric excess noise level of 0.11, while relatively low, ideally should be zero. Astrometric excess noise can indicate modeling errors in either the excess noise associated with the source or the attitude noise. These hints of astrometric variations must be interpreted with caution and drawing conclusions is beyond the scope of this paper. Gaia DR4 will enable the determination of WASP-132\,d's true mass and its relative orbital inclination with respect to the two inner transiting planets. We note that WASP-132\,d would need to have an orbital inclination larger than 66 degrees from an edge-on orbit to have a mass greater than 13 $M_{\rm{Jup}}$.

\subsection{Internal characterization of the giant planet WASP-132\,b}

\begin{figure}
  \centering
\includegraphics[width=0.5\textwidth]{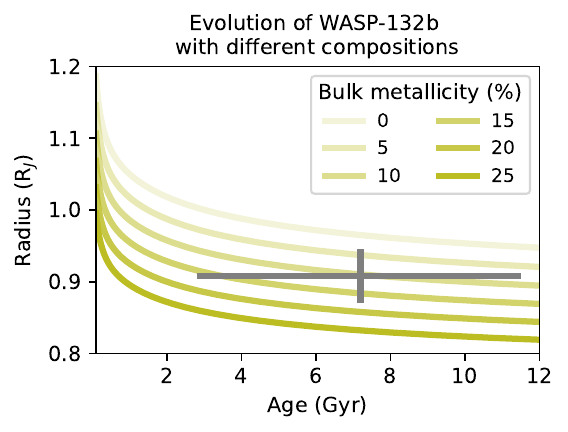}
 \caption{WASP-132\,b's radius as a function of time. The green lines display the cooling of WASP-132\,b assuming different bulk metallicity values between zero and 25\%. The gray error bars show the measured values. The planet must be enriched in heavy elements to match the observations.}
  \label{fig:wasp132b_evolution}
\end{figure}

The available measurements of the mass and the radius of the planet WASP-132\,b make it possible to infer its bulk heavy-element content with giant planet evolution models \citep[e.g.,][]{Miller2011,Thorngren2016,Mueller2023a}. This is an important piece of information since it can be compared to predictions from formation models and therefore be used as an additional constraint \citep[e.g.,][]{Guillot2006,Johansen2017,Hasegawa2018}. With a $T_{eq} = 682$ K, WASP-132\,b is significantly cooler than strongly irradiated hot Jupiters. This makes the inferred composition less uncertain since the planet should not be inflated \citep[e.g.,][]{Fortney2021}. Here, we used the giant planet evolution models from \citet{Mueller2021} to calculate the planet's cooling. Figure \ref{fig:wasp132b_evolution} shows the radius as a function of time assuming that the planet has a bulk metallicity between zero and and 25\%. Comparing it to the observed radius, it is clear that the planet is not inflated compared to predictions from the model, and that it must be enriched with heavy elements.

\begin{figure}
  \centering
\includegraphics[width=0.5\textwidth]{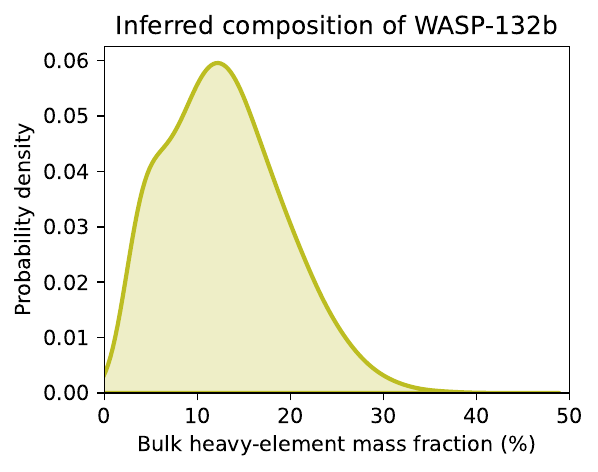}
 \caption{Posterior distribution of the inferred heavy-element mass fraction of WASP-132b. The distribution is approximately Gaussian, with a mean of $\mu_{Z} = 13 \%$ and a standard deviation of $\sigma_Z = 6 \%$}
  \label{fig:wasp132b_metallicity}
\end{figure}

To quantify the enrichment, we inferred the planet's bulk heavy-element mass fraction $Z$ in a Monte-Carlo fashion \citep[see e.g.,][for a review]{Mueller2023b}: we create sample planets by drawing from the observed planetary mass, radius, and age distributions. For each sample, we calculate the evolution for a range of bulk metallicities to find which value would match the sampled radius at the right age. By repeating this process, the posterior distribution of the planet's heavy-element mass can be estimated. The result is shown in Fig. \ref{fig:wasp132b_metallicity}. The posterior is roughly Gaussian, with a mean of $\mu_{Z} = 13 \%$ and a standard deviation of $\sigma_Z = 6 \%$. This yields a heavy-element mass of $M_z\approx17\,M_\oplus$, which is similar in magnitude to what would be expected from the critical core mass in core-accretion models \citep[e.g.,][]{Helled2014}. For comparison, if the planet had stellar metallicity, its heavy-element mass would be $M_Z \approx 2.6 M_\oplus$.

\subsection{Internal characterization of WASP-132\,c}

Figure \ref{fig:mr_diagram} shows a Mass-Radius diagram highlighting multi-planetary systems with precise densities ($\sigma_{M}/M$ $\leq$ 25$\%$ and $\sigma_{R}/R$ $\leq$8$\%$) from the PlanetS catalog{\footnote{https://dace.unige.ch/exoplanets/}} (\citealt{Otegi_2020,Parc2024}), and in particular systems composed of a hot or warm Jupiter (P$<$100 days) and a nearby small planet (R\,$<$\,4\,$R_\oplus$). Figure \ref{fig:mr_diagram} clearly shows that WASP-132\,c sits slightly above the Earth-like composition line, suggesting a refractory-rich composition. To probe the scope of interior structures that are consistent with the observed constraints, we use the four layer model of \citet{Dorn_2017}. It includes an iron core, a rocky mantle, a water layer, and a hydrogen-helium atmosphere. Using a nested sampling algorithm \citep{Buchner_2014}, we vary the layer masses, the elemental ratios of Mg/Si and Fe/Si in the mantle and the age of the planet, taking the stellar values for the priors. In order to put upper bounds on the water and atmospheric mass fractions we ran a model assuming no atmosphere (no-atmosphere model) and a model assuming no water layer (no-water model), respectively. Additionally, we ran a model where we put no constraints on the layer masses (free model). Our results are listed in Table \ref{tab:interior_nested_sampling}.

\begin{table}
    \tiny
    \centering
    \caption{WASP-132\,c interior structure nested sampling results.}
    \begin{tabular}{lcccc}
        \hline
        \hline
        \noalign{\smallskip}
         Model & $M_\mathrm{core}/M_p$ & $M_\mathrm{mantle}/M_p$ & $M_\mathrm{water}/M_p$ & $\log (M_\mathrm{atm}/M_p)$\\
         \noalign{\smallskip}
        \hline
        \noalign{\smallskip}
        free & $0.53 \pm 0.23$ & $0.38 \pm 0.24$ & $0.09 \pm 0.06$ & $-8.54 \pm 1.85$ \\
        no-atmosphere & $0.50 \pm 0.23$ & $0.38 \pm 0.24$ & $0.11 \pm 0.06$ & N/A \\
        no-water & $0.47 \pm 0.26$ & $0.53 \pm 0.26$ & $0.00$ & $-5.08 \pm 1.17$ \\
         \hline
    \end{tabular}
    \label{tab:interior_nested_sampling}
\end{table}

All nested sampling models favor a refractory-rich composition dominated by metals and silicates and a negligible atmosphere. It should be noted, however, that the assumed internal structure is very simple. In reality, the internal structure of planets can be more complex, including and enrichment of the atmosphere with heavy elements and the hydration of the mantle and/or core. Accounting for more sophisticated structure could affect the inferred results. However, our main conclusions that WASP-132c is an exoplanet dominated by refractory material are unlikely to change.

\subsection{WASP-132: A rare multi-planet system}

\begin{figure}
    \centering
    \includegraphics[width=0.5\textwidth]{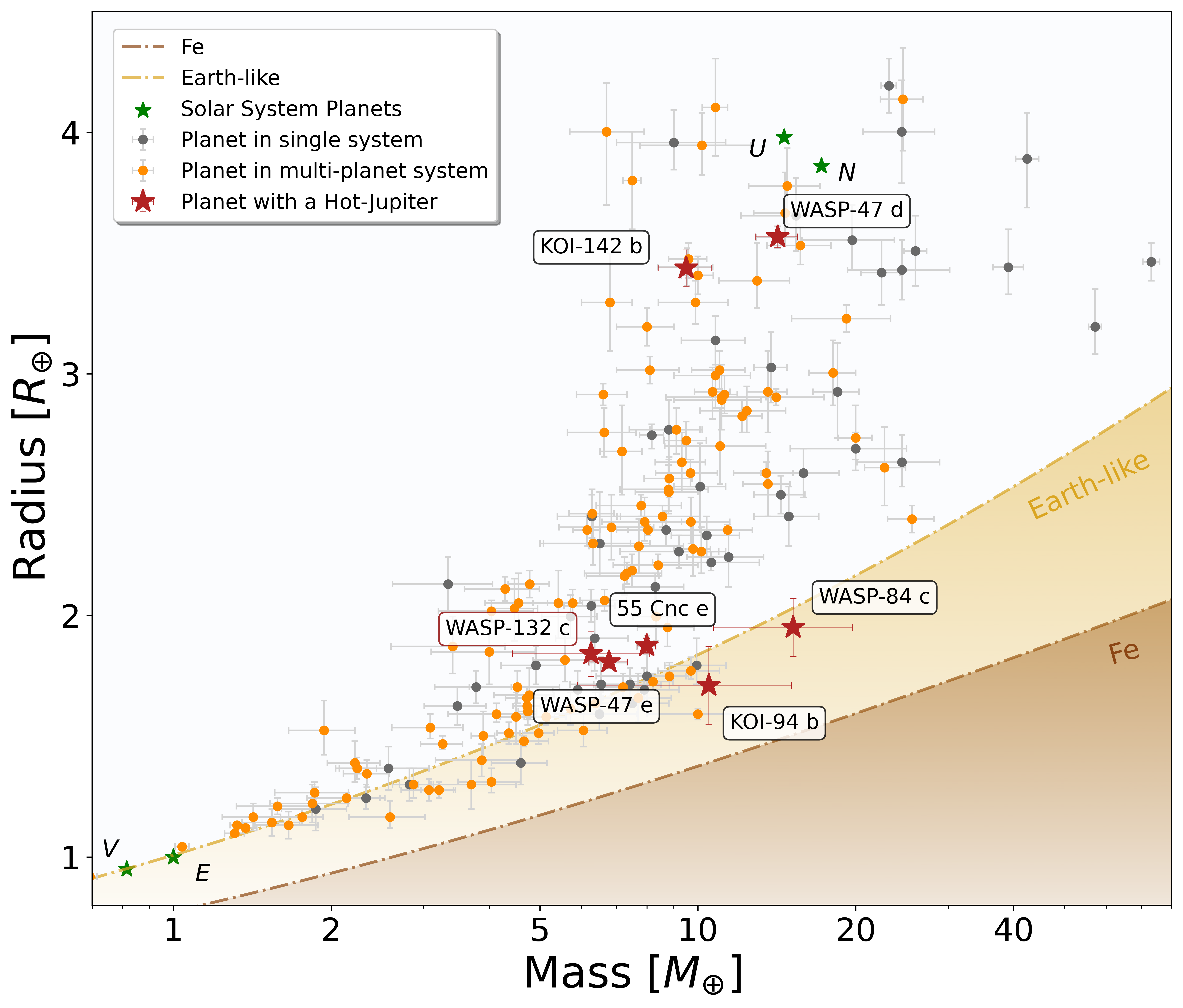}
    \caption{Mass-radius diagram of small exoplanets with precise densities ($\sigma_{M}/M$ $\leq$ 25$\%$ and $\sigma_{R}/R$ $\leq$8$\%$) from the PlanetS catalog (\citealt{Otegi_2020,Parc2024}). The red stars correspond to known small planets (R\,$<$\,4\,$R_\oplus$), with precise radius and mass measurements, that have a hot or warm Jupiter companion (R\,$>$\,0.8 R$_{\rm{Jup}}$). Orange dots are planets in multi-planet system of all kinds. Gray dots are planets in single-planet system. Composition lines are displayed for pure-iron (brown) and Earth-like planets (light-brown).}
    \label{fig:mr_diagram}
\end{figure}

As noted, Fig. \ref{fig:mr_diagram} shows a mass-radius diagram highlighting multi-planetary systems with precise densities from the PlanetS catalog. Among the 127 planets in 80 systems, only six have a hot or warm Jupiter (P$<$100 days) and a nearby small planet (R\,$<$\,4\,$R_\oplus$): WASP-132, WASP-47 (\citealt{Bryant2022}), 55 Cnc (\citealt{Bourrier2018}), KOI-94 (\citealt{Weiss2013}), WASP-84 (\citealt{Maciejewski2023}) and KOI-142 (\citealt{Weiss2020}). This once again demonstrates the rare contribution of this system, and the mass measurement of WASP-132\,c, to a very small existing population. 

The small planets with accompanying hot Jupiters consist of five super-Earths and two sub-Neptunes. Considering the sub-Neptunes, WASP-47\,d is located at a longer period than its hot Jupiter companion, while the super-Earth WASP-47\,e is inside the hot Jupiter. KOI-142\,b orbits with a much longer period than the other planets in this diagram (P $=$ 10.92 days), explaining its larger size due to lower irradiation. Its associated Jupiter is also less "hot." 

The super-Earths with hot or warm Jupiters are on the more massive end of this planet type and are likely composed of refractory elements. This may reveal that we are biased to detect the heaviest super-Earths in the presence of a hot Jupiter and lighter inner rocky planets in this kind of system may exist. Smaller planets like Kepler-730\,c (R\,=\,1.57\,$\pm$\,0.13\,$R_\oplus$; \citealt{Canas2019}), which have no mass measurement, could be lighter. The lack of a similar system between the super-Earths and the sub-Neptunes could be intimately linked to their short orbital period and therefore to the irradiation received, stripping these planets of their atmosphere. Hot Jupiters then play an important role behind, limiting an internal planet's period.

\section{Conclusion} \label{sec:conclusion}

We report refined bulk measurements of the WASP-132 system. The 7.1-day hot Jupiter WASP-132\,b was first discovered by WASP \citep{Hellier2017} and an inner 1.01-day super-Earth was subsequently discovered by TESS \citep{Hord2022}. Here we also report the discovery of an 2.7 AU outer giant planet and long-term linear trend in the RVs suggestive of another outer companion. Using over nine years of CORALIE observations and highly sampled HARPS data, we determined the masses of the planets from smallest to largest orbital period to be M$_{\rm{p}}$\,=\,$6.26^{+1.84}_{-1.83}$\,$M_{\oplus}$, M$_{\rm{p}}$\,=\,$0.428^{+0.015}_{-0.015}$\,$M_{\rm{Jup}}$, and M$_{\rm{p}}\sin{i}$\,=\,$5.16^{+0.52}_{-0.52}$\,$M_{\rm{Jup}}$, respectively. Using TESS and CHEOPS photometry data, we measured the radii of the two inner transiting planets to be $1.841^{+0.094}_{-0.093}$\,$R_{\oplus}$ and $0.901^{+0.038}_{-0.038}$\,$R_{\rm{Jup}}$. We also performed an independent stellar analysis to find M$_{\star}$\,=\,$0.789\,\pm\,0.039$\,$M_{\odot}$ and R$_{\star}$\,=\,$0.758\,\pm\,0.032$\,$R_{\odot}$. We find that the hot Jupiter WASP-132\,b is not inflated but enriched with heavy elements. For the super-Earth WASP-132\,c, our internal structure modeling favors a refractory-rich composition that is dominated by metals and silicates and a negligible atmosphere. We find that both the inner 1-day super-Earth and hot Jupiter are likely in circular orbits with eccentricities of $\sim$0. Given the presence of a nearby inner planet and its low eccentricity, the hot Jupiter WASP-132\,b likely migrated through quieter mechanisms than high-eccentricity migration, making it a rare contribution to the current planet population. 

\section{Data availability}

A table of the reduced CHEOPS photometric observations is available in a machine-readable format at The Strasbourg astronomical Data Center (CDS) via anonymous ftp to cdsarc.u-strasbg.fr (130.79.128.5) or via \href{http://cdsweb.u-strasbg.fr/cgi-bin/qcat?J/A+A/}{http://cdsweb.u-strasbg.fr/cgi-bin/qcat?J/A+A/}. The reduced TESS photometric observations are available in a machine-readable format at the CDS. The CORALIE RVs and all activity indicator data are available in a machine-readable format at the CDS. The HARPS RVs and all activity indicator data are available in a machine-readable format at the CDS. Figures mentioned in the text as supplementary material are published on Zenodo at \href{https://doi.org/10.5281/zenodo.14271139}{https://doi.org/10.5281/zenodo.14271139}.

\begin{acknowledgements}
We thank the Swiss National Science Foundation (SNSF) and the Geneva University for their continuous support to our planet low-mass companion search programs. This work was carried out in the frame of the Swiss National Centre for Competence in Research (NCCR) $PlanetS$ supported by the SNSF including grants 51NF40\_182901 and 51NF40\_205606. This work used computations that were performed at the University of Geneva on the "Yggdrasil" High Performance Computing (HPC) clusters. This publication makes use of The Data \& Analysis Center for Exoplanets (DACE), which is a facility based at the University of Geneva (CH) dedicated to extrasolar planet data visualization, exchange, and analysis. DACE is a platform of NCCR $PlanetS$ and is available at https://dace.unige.ch. This paper includes data collected by the TESS mission. Funding for the TESS mission is provided by the NASA Explorer Program. We acknowledge the use of public TESS data from pipelines at the TESS Science Office and at the TESS Science Processing Operations Center. Resources supporting this work were provided by the NASA High-End Computing (HEC) Program through the NASA Advanced Supercomputing (NAS) Division at Ames Research Center for the production of the SPOC data products. V.A. acknowledges the support by FCT - Fundação para a Ciência e a Tecnologia through national funds (grants: UIDB/04434/2020, UIDP/04434/2020, and 2022.06962.PTDC). DJA is supported by UKRI through the STFC (ST/R00384X/1) and EPSRC (EP/X027562/1). NCS acknowledges the funding by the European Union (ERC, FIERCE, 101052347). Views and opinions expressed are however those of the author(s) only and do not necessarily reflect those of the European Union or the European Research Council. Neither the European Union nor the granting authority can be held responsible for them. This work was supported by FCT - Fundação para a Ci\^encia e a Tecnologia through national funds and by FEDER through COMPETE2020 - Programa Operacional Competitividade e Internacionalização by these grants: UIDB/04434/2020; UIDP/04434/2020. N.G. thanks Daniel Thomas Watson for his pertinacious and adamantine support. 

\end{acknowledgements}

%
%

\bibliographystyle{aa}
\bibliography{bib}

\end{document}